\documentclass[10pt]{article}
\usepackage[T1]{fontenc}
\usepackage[utf8]{inputenc}
\usepackage{amsmath,amssymb,amsthm,mathtools}
\usepackage{preprint}
\usepackage{graphicx,booktabs,tabularx,multirow,array,makecell,adjustbox}
\usepackage{caption,subcaption}
\usepackage{placeins,stfloats,balance,enumitem}
\usepackage[numbers,sort&compress]{natbib}
\usepackage[hidelinks]{hyperref}
\usepackage{microtype}
\usepackage{xurl,etoolbox}
\providecommand{\Description}[2][]{}
\AtBeginEnvironment{equation}{\small}
\AtBeginEnvironment{align}{\small}
\makeatletter
\def\verbatim@font{\normalfont\ttfamily\scriptsize}
\makeatother
\date{}
\newenvironment{prompttext}{\begin{list}{}{\setlength{\leftmargin}{1.5em}\setlength{\rightmargin}{1.5em}\setlength{\topsep}{2.5pt}\setlength{\partopsep}{0pt}}\item\relax\footnotesize}{\end{list}}
\newcommand{\slot}[1]{\textit{[#1]}}
\title{Reversing Polarization through Tolerance-Aware Recommendation and Adaptive Diversification}
\author{Zhanzhan Zhao\textsuperscript{1,2, *}\quad Shuqi Zhao\textsuperscript{2}\quad Zichen Fan\textsuperscript{2}\quad Dana Randall\textsuperscript{3}}
\preprintaffiliations{\textsuperscript{1}School of Humanities and Social Science, The Chinese University of Hong Kong, Shenzhen, Shenzhen, China\\[2pt]
\textsuperscript{2}School of Artificial Intelligence, The Chinese University of Hong Kong, Shenzhen, Shenzhen, China\\[2pt]
\textsuperscript{3}School of Computer Science, Georgia Institute of Technology, Atlanta, USA\\[2pt]
\textsuperscript{*}Corresponding author: Zhanzhan Zhao (\href{mailto:zhanzhanzhao@cuhk.edu.cn}{zhanzhanzhao@cuhk.edu.cn})}
\begin{document}
\maketitle
\begin{abstract}
Recommender systems face a tension between broadening exposure to diverse viewpoints and avoiding interactions that reinforce extreme positions. This tension is especially consequential in already polarized populations, where limiting disagreement may reduce backlash while preserving divisions between opposing groups. Here we investigate whether recommendation strategies can reverse polarization rather than merely slow its development. Building on the Attraction--Repulsion Model of opinion dynamics, we introduce two complementary design principles: Tolerance-aware recommendation, which seeks to learn each individual's tolerance for disagreement and reduces exposure to viewpoints beyond that tolerance, and adaptive diversification, which broadens exposure for agents with more extreme opinions while retaining stronger personalization for moderates. Agent-based experiments show that highly personalized recommendations intensify polarization under polarized initial conditions. Tolerance awareness alone primarily slows polarization, whereas combining it with adaptive diversification can reduce polarization below its initial level and, under suitable conditions, restore consensus. Additional experiments incorporating persistent extreme influencers and a spiral-of-silence mechanism show that the combined strategy can mitigate polarization under social pressure and selective participation. To examine whether these findings extend beyond predefined opinion-update rules, we repeat the experiments on a simulated social platform populated by 100 LLM-driven agents prompted to move toward opinions similar to their own and away from those that differ substantially. Across three runs, the combined strategy is again the only recommendation policy tested that reverses polarization in every run. These findings distinguish the prevention of polarization from its reversal and identify tolerance-aware, adaptive exposure as a promising recommendation design principle, providing a mechanism-based foundation for reconnecting divided populations while accounting for users' responses to disagreement.
\end{abstract}
\keywords{Recommender systems, Polarization, Opinion dynamics, Attraction--Repulsion Model, Agent-based modeling, Spiral of silence, Large language models, LLM-based agents}
\vspace{12pt}

\section{Introduction}

Extreme ideological polarization can undermine democratic compromise by making mutual understanding across ideological groups increasingly difficult~\cite{sunstein2018republic}. Algorithmic recommender systems, which tailor content and social connections to users' preferences, have therefore attracted considerable attention as potential drivers of polarization through selective exposure and ideologically segregated information environments~\cite{brugnoli2019recursive,flaxman2016filter,cinelli2021echo}. Yet efforts to counteract polarization often turn to diversity-enhancing algorithms that simply increase exposure to divergent content, overlooking a crucial psychological constraint: when individuals encounter views that fall beyond their tolerance threshold, the result may be not accommodation but repulsion, driving them toward more extreme positions~\cite{axelrod1997dissemination,bail2018exposure,noorazar2020classical,mas2013differentiation}.

This repulsion-driven dynamic is precisely captured by the Attraction--Repulsion Model (ARM) introduced by Axelrod et al.~\cite{axelrod2021preventing}. The ARM is a minimal agent-based model with two simple rules. The first rule governs the interaction: The probability that two agents interact decays exponentially with their opinion distance, scaled by an exposure parameter that determines how quickly the probability declines as distance increases. Once an interaction occurs, its effect depends on a common tolerance threshold. If their distance falls within this threshold, the active agent moves toward the passive agent (attraction); if their distance exceeds it, the active agent moves away (repulsion). Using this model, prior work showed that under normally distributed initial opinions, ultra-personalized recommendations (low exposure parameter) can mitigate polarization by limiting repulsive cross-cutting interactions across a range of tolerance levels~\cite{axelrod2021preventing}. However, these results were obtained under a unimodal, symmetric initial distribution. Whether the same mitigating effect holds when the population is already polarized --- and whether recommender strategies could be redesigned to reverse, not merely slow, polarization --- remains unknown.

To address these gaps, we extend the ARM in three directions. First, we introduce tolerance-aware recommendation into the interaction rule: the probability of interaction depends not only on the distance and exposure parameter, but also on the tolerance threshold. When the distance between two agents falls within the tolerance threshold, their interaction probability follows the standard exposure-based rule; when it exceeds the tolerance, a penalty factor reduces that probability. This mechanism aims to reduce repulsive interactions without eliminating all cross-cutting exposure. Second, we propose adaptive diversification: the degree of personalization is made a function of an agent's distance from the reference position. Agents near the reference position maintain high personalization, while those at the extremes receive more diverse recommendations. Third, we incorporate the spiral of silence and extreme influencers, allowing us to examine whether our proposed strategies remain effective under social pressure --- conditions known to intensify polarization~\cite{noelleneumann1991theory,lazarsfeld1968people,huszar2022algorithmic}.

Using agent-based simulations, our extended ARM shows that recommendation strategies combining tolerance-awareness with adaptive diversification can reverse an ongoing polarizing trend, even when the population is already polarized. In contrast, personalized recommendations alone --- based solely on opinion distance and exposure parameter --- fail under these conditions. This mitigating effect remains robust even when the spiral of silence and extreme influencers amplify polarization.

Like other agent-based models, these results depend on behavioral rules specified by the modeler. We therefore also test them on a simulated platform whose 100 users are agents driven by a large language model (LLM)~\cite{gao2024large}. We prompt the LLM agents with the attraction--repulsion premise, motivated by psychological evidence that people evaluate belief-consistent information more favorably and may respond to conflicting information by reinforcing their existing positions \cite{lord1979biased,taber2006motivated}. Specifying this premise keeps the behavioral assumptions consistent across all LLM experiments while allowing the LLM to generate individual responses without a fixed numerical update rule. Under these conditions, ultra-personalized and tolerance-aware feeds leave the two camps divided, while the combined design is the only policy tested that reverses polarization in all three runs. Excessive diversification widens the divide again, and under social pressure the combined design still lowers polarization, though less than in the rule-based model.

These findings suggest that when societies are already divided, reducing exposure based solely on similarity is insufficient; incorporating tolerance into the interaction rule offers a promising mechanism for reversing polarization. While our simplified model does not capture the full complexity of real-world systems, it provides insights into how exposure can be calibrated to both individual tolerance thresholds and the current distribution of opinions.

\subsection{Related Work}
\label{sec:related-work}

Research on recommendation and polarization emphasizes the interaction between exposure and behavioral responses to disagreement. Axelrod et al.~\cite{axelrod2021preventing} distinguish encounter probabilities from opinion updates in the Attraction--Repulsion Model, showing how tolerance and exposure jointly influence polarization. Santos et al.~\cite{santos2021link} similarly demonstrate that the consequences of link recommendation depend on network structure and individuals' responses to outgroup contact. These models caution against treating greater exposure diversity as inherently beneficial. Empirically, Bail et al.~\cite{bail2018exposure} find that Republican participants exposed to opposing political views became more conservative, illustrating that cross-cutting exposure can produce backlash in some settings. Together, these studies motivate recommendation policies that account for responses to disagreement alongside opinion similarity.

Prior research also establishes that interventions can reduce existing polarization. Currin et al.~\cite{currin2022depolarization} introduce random dynamical nudges based on sampled population opinions and demonstrate depolarization of established echo chambers in two opinion-dynamics models. Piccardi et al.~\cite{piccardi2025reranking} provide complementary field evidence that reranking exposure to posts expressing antidemocratic attitudes and partisan animosity can alter affective polarization. Their outcome concerns hostility toward political opponents, whereas our model examines dispersion in opinion positions. Our study contributes to this literature by examining tolerance-aware interaction selection together with opinion-dependent exposure within the ARM. We investigate when their combination moves beyond slowing polarization to reversing an existing division, including settings where low tolerance or excessive diversification undermines the benefit.

LLM-based social simulation offers an additional approach to examining recommendation policies. T{\"o}rnberg et al.~\cite{tornberg2023simulating} compare alternative news-feed algorithms and find that a bridging feed promotes more constructive conversation across political divides. Donkers and Ziegler~\cite{donkers2025understanding} investigate human responses to polarized environments populated by LLM agents. However, Chuang et al.~\cite{chuang2024simulating} identify an inherent tendency toward consensus in LLM opinion simulations, while Ju et al.~\cite{ju2024sense} demonstrate sensitivity to minor prompt variations. These findings motivate explicit behavioral checks and comparison with reference dynamics. We use LLM agents to examine whether the qualitative policy comparison persists when opinion updates are generated under a shared, explicitly prompted attraction--repulsion premise.

\section{Model and Design Principles}
\label{sec:model}

Our framework extends the Attraction--Repulsion Model (ARM) by incorporating three mechanisms: tolerance-aware exposure, adaptive diversification, and social pressure through the spiral of silence together with influencer agents. Jointly, these components allow us to systematically examine how different recommendation strategies shape polarization dynamics within a simulated population.

\begin{figure}[t]
  \centering
  \begin{subfigure}{0.74\linewidth}
    \centering
    \includegraphics[width=\linewidth]{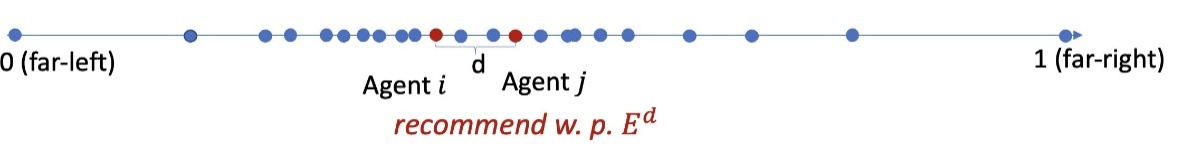}
    \caption{Interaction probability decays exponentially with opinion distance, parameterized by $E$.}
    \label{fig:schematic-exposure}
  \end{subfigure}

  \medskip
  \begin{subfigure}{0.74\linewidth}
    \centering
    \includegraphics[width=\linewidth]{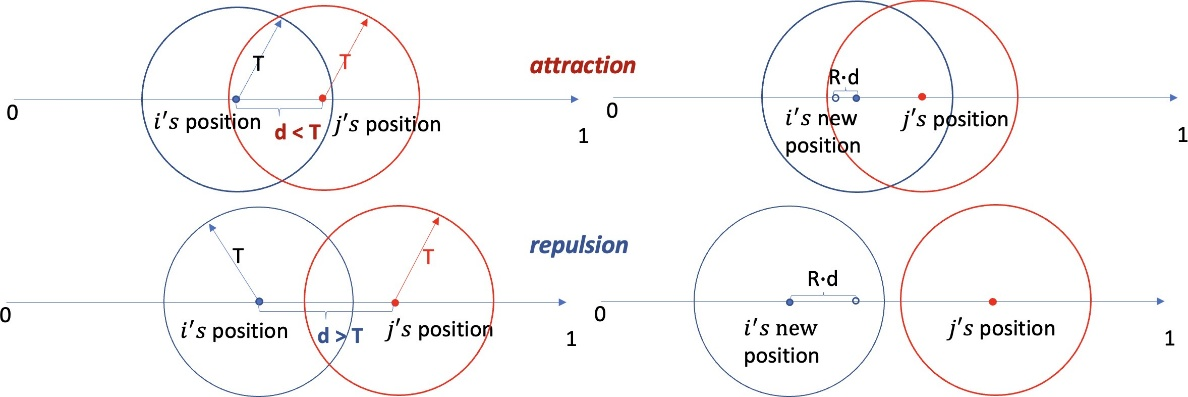}
    \caption{Attraction--repulsion rule updates opinions based on tolerance threshold $T$.}
    \label{fig:schematic-rule}
  \end{subfigure}
  \caption{Schematic illustration of the extended Attraction--Repulsion Model (ARM).}
  \Description{Panel (a) shows agents as points on the opinion line from 0 (far-left) to 1 (far-right); agent i is recommended agent j at distance d with probability E to the power d. Panel (b) shows the attraction case, where the distance is below T and i moves R times d toward j, and the repulsion case, where the distance exceeds T and i moves R times d away from j.}
  \label{fig:schematic}
\end{figure}

Consider a population of $N$ agents, where each agent $i$ is assigned a continuous position $x_i \in [0,1]$ representing its location in the opinion space. At each discrete step $t$, an ordered pair $(i,j)$ of distinct agents is drawn uniformly at random. Agent $i$ is active and agent $j$ is passive; the pair interacts with the probability specified below, whereas a rejected attempt leaves all opinions unchanged and still counts as one step. With the opinion distance between agents $i$ and $j$ defined as $d_{ij}(t) = |x_i(t) - x_j(t)|$, the baseline interaction probability is governed by:
\begin{equation}
  P_{ij}(t) = E^{d_{ij}(t)}
  \label{eq:exposure}
\end{equation}
where $E \in [0,1]$ controls the level of exposure. In our formulation, smaller values of $E$ correspond to stronger personalization by assigning greater relative weight to interactions among agents with smaller opinion distances, whereas larger values of $E$ allow broader exposure across the opinion space (see Figure~\ref{fig:schematic}).

When an interaction occurs, only the active agent $i$ updates its opinion; agent $j$ remains unchanged. All agents share a common tolerance threshold $T$. If $d_{ij} \le T$, agent $i$ moves a fraction $R$ of the distance toward $j$ (attraction); if $d_{ij} > T$, it moves the same fraction away from $j$ (repulsion). Before clipping, the update is
\begin{equation}
  \widetilde{x}_i(t+1) =
  \begin{cases}
    x_i(t) + R[x_j(t) - x_i(t)] & d_{ij}(t) \le T,\\
    x_i(t) - R[x_j(t) - x_i(t)] & d_{ij}(t) > T.
  \end{cases}
  \label{eq:update}
\end{equation}
Clipping restricts the updated opinion to the valid interval: $x_i(t+1) = \max\{0, \min\{1, \widetilde{x}_i(t+1)\}\}$. Responsiveness is governed by $R \in [0,1]$, which scales the magnitude of adjustment, and polarization is quantified as the population variance of opinions, $\mathrm{Var}(x_i)$.

\begin{figure}[t]
  \centering
  \begin{subfigure}[t]{0.38\linewidth}
    \centering
    \includegraphics[width=\linewidth]{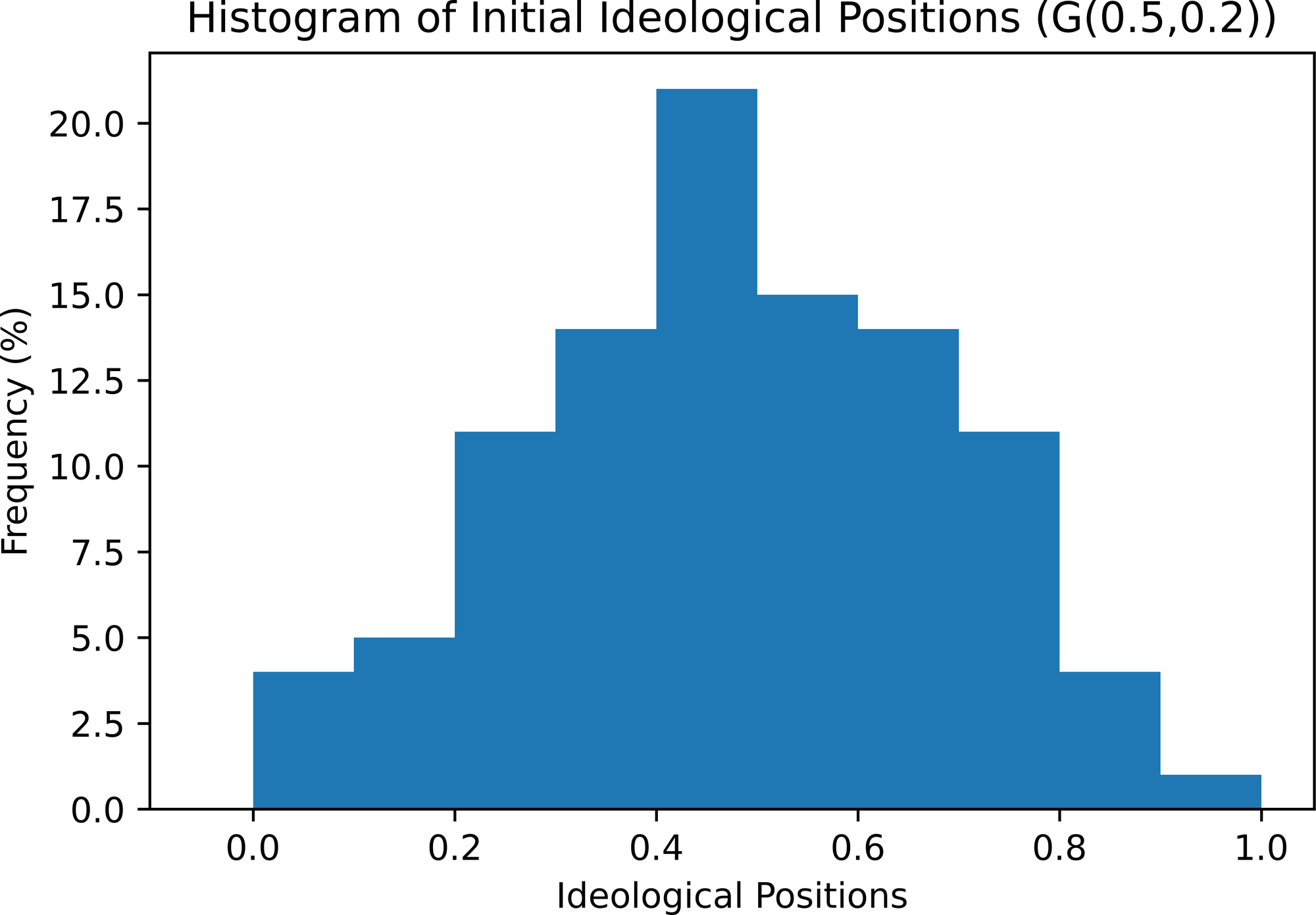}
    \caption{Truncated Gaussian distribution (sample variance $\approx 0.035$; SD $0.2$ before truncation).}
    \label{fig:init-gaussian}
  \end{subfigure}\hspace{0.035\linewidth}
  \begin{subfigure}[t]{0.38\linewidth}
    \centering
    \includegraphics[width=\linewidth]{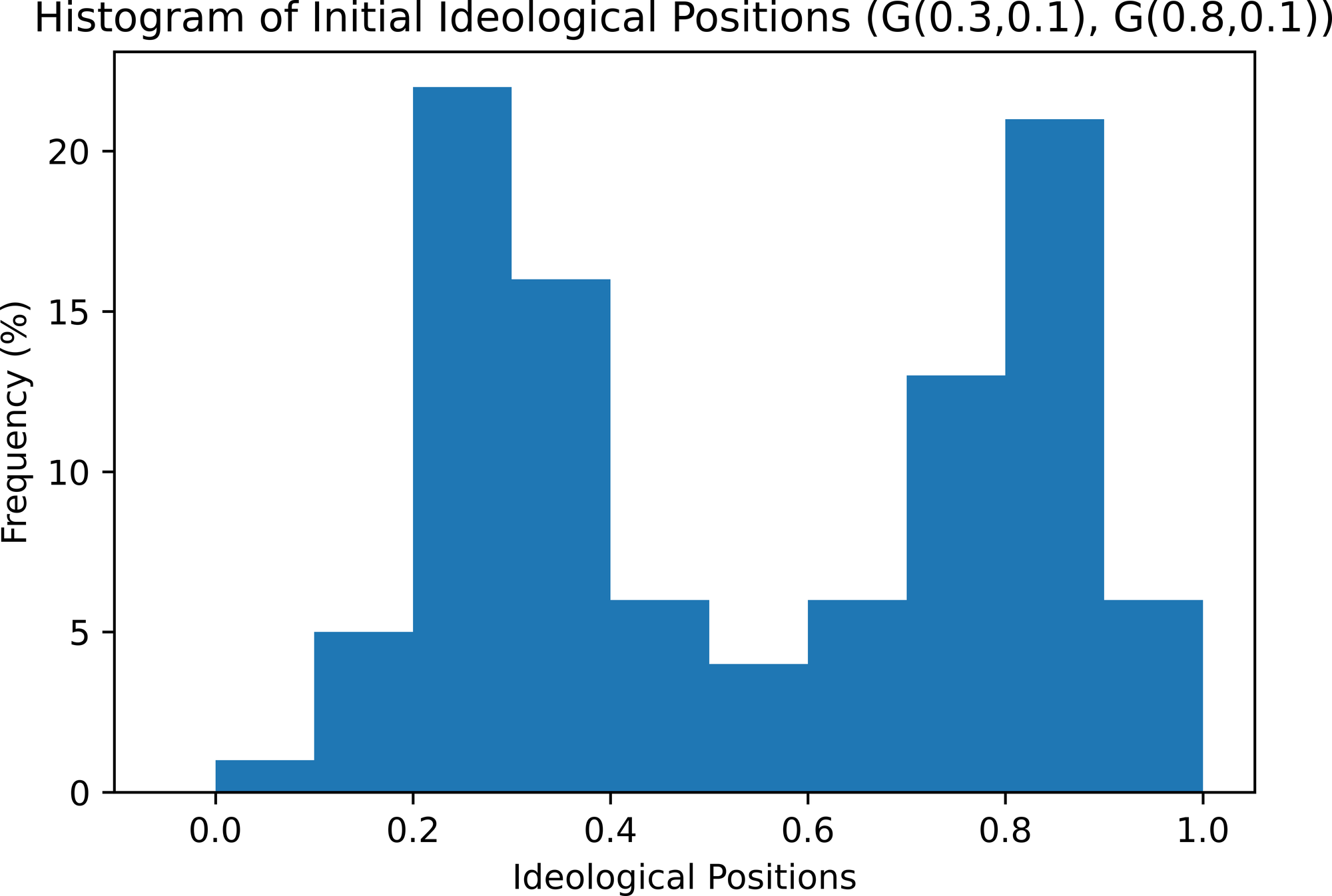}
    \caption{Polarized bimodal distribution (variance = 0.074).}
    \label{fig:init-bimodal}
  \end{subfigure}
  \caption{Initial ideological configurations used in simulations.}
  \Description{Two histograms of initial ideological positions on [0, 1]. The left one is unimodal, peaking near 0.45. The right one is bimodal, with peaks near 0.25 and 0.85.}
  \label{fig:init}
\end{figure}

\subsection{Tolerance-Aware Recommendation}

In the baseline ARM, the tolerance threshold determines whether a selected interaction results in attraction or repulsion. Here, we extend its role by allowing the same threshold to additionally influence interaction selection while preserving the personalized exposure mechanism. Consequently, personalization continues to determine the baseline interaction probability, whereas tolerance introduces an additional adjustment for interactions between agents whose opinion distance exceeds the tolerance threshold.

Specifically, interactions between agents whose opinion distance falls within the tolerance threshold follow the personalized exposure rule defined above. Beyond that threshold, the personalized interaction probability is attenuated by a penalty factor controlled by the tolerance-awareness parameter $E'$:
\begin{equation}
  P_{ij}(t) =
  \begin{cases}
    E^{d_{ij}(t)} & \text{if } d_{ij}(t) \le T\\[2pt]
    \dfrac{E^{d_{ij}(t)}}{E'} & \text{if } d_{ij}(t) > T
  \end{cases}
  \label{eq:tolerance-aware}
\end{equation}
where $E' \ge 1$ controls the strength of the tolerance-aware adjustment. Setting $E' = 1$ recovers the personalized exposure rule defined in the baseline model. Increasing $E'$ progressively decreases the probability of interactions between agents whose opinion distance exceeds the tolerance threshold, while leaving interactions within the tolerance range unchanged.

Personalization thus remains the primary mechanism governing interaction selection, while tolerance serves a dual role in regulating opinion updates and influencing the likelihood that interactions occur. Throughout the simulations, $E'$ varies from 1 (no tolerance-aware adjustment) to 100 (strong tolerance-aware adjustment).

\begin{figure}[t]
  \centering
  \begin{subfigure}[t]{0.44\linewidth}
    \centering
    \includegraphics[width=\linewidth]{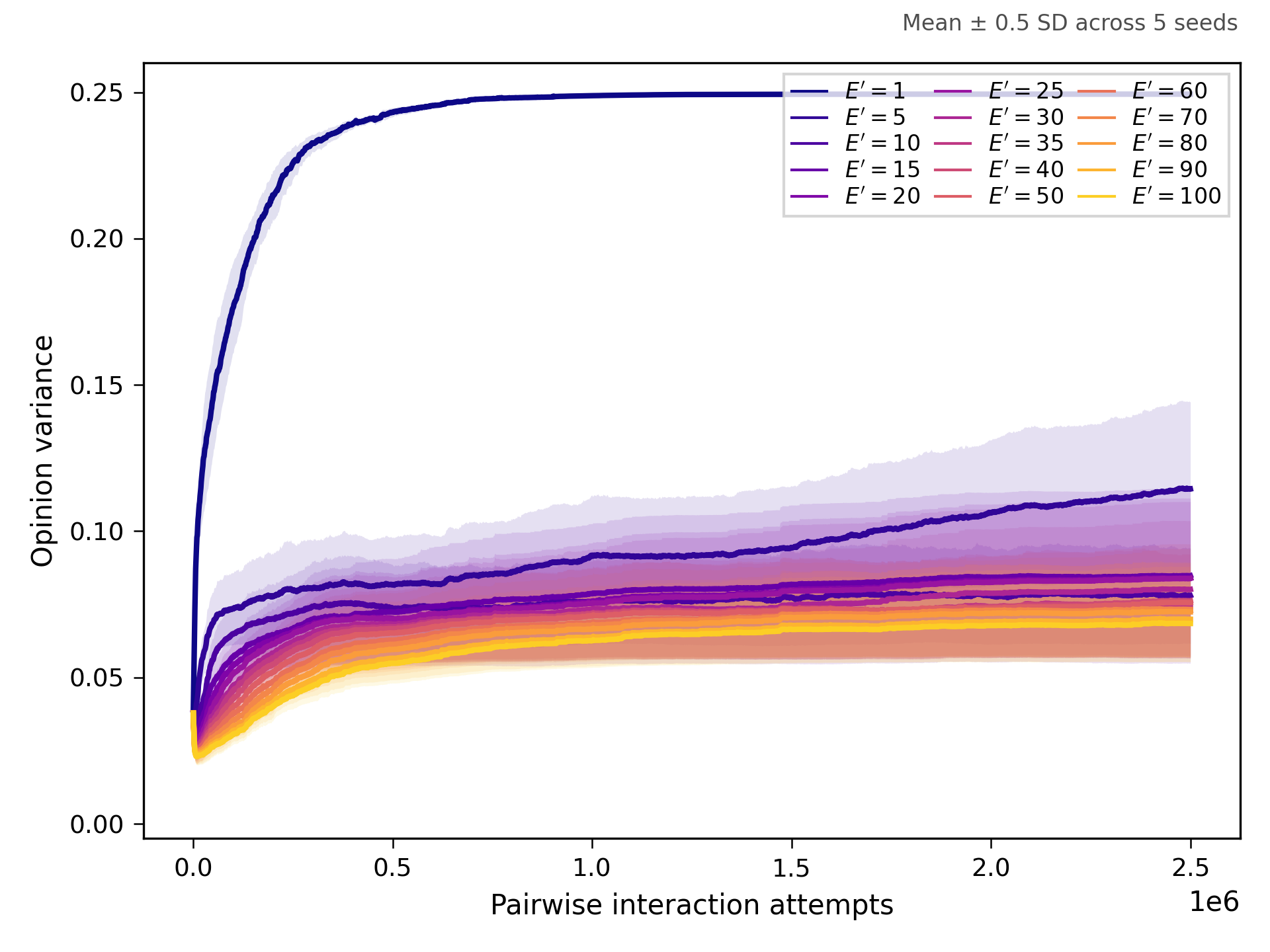}
    \caption{Polarization dynamics under varying tolerance-awareness parameter $E'$ ($E = 0.5^{10}$, $T = 0.2$, normal initialization).}
    \label{fig:sweep-Eprime}
  \end{subfigure}\hspace{0.035\linewidth}
  \begin{subfigure}[t]{0.44\linewidth}
    \centering
    \includegraphics[width=\linewidth]{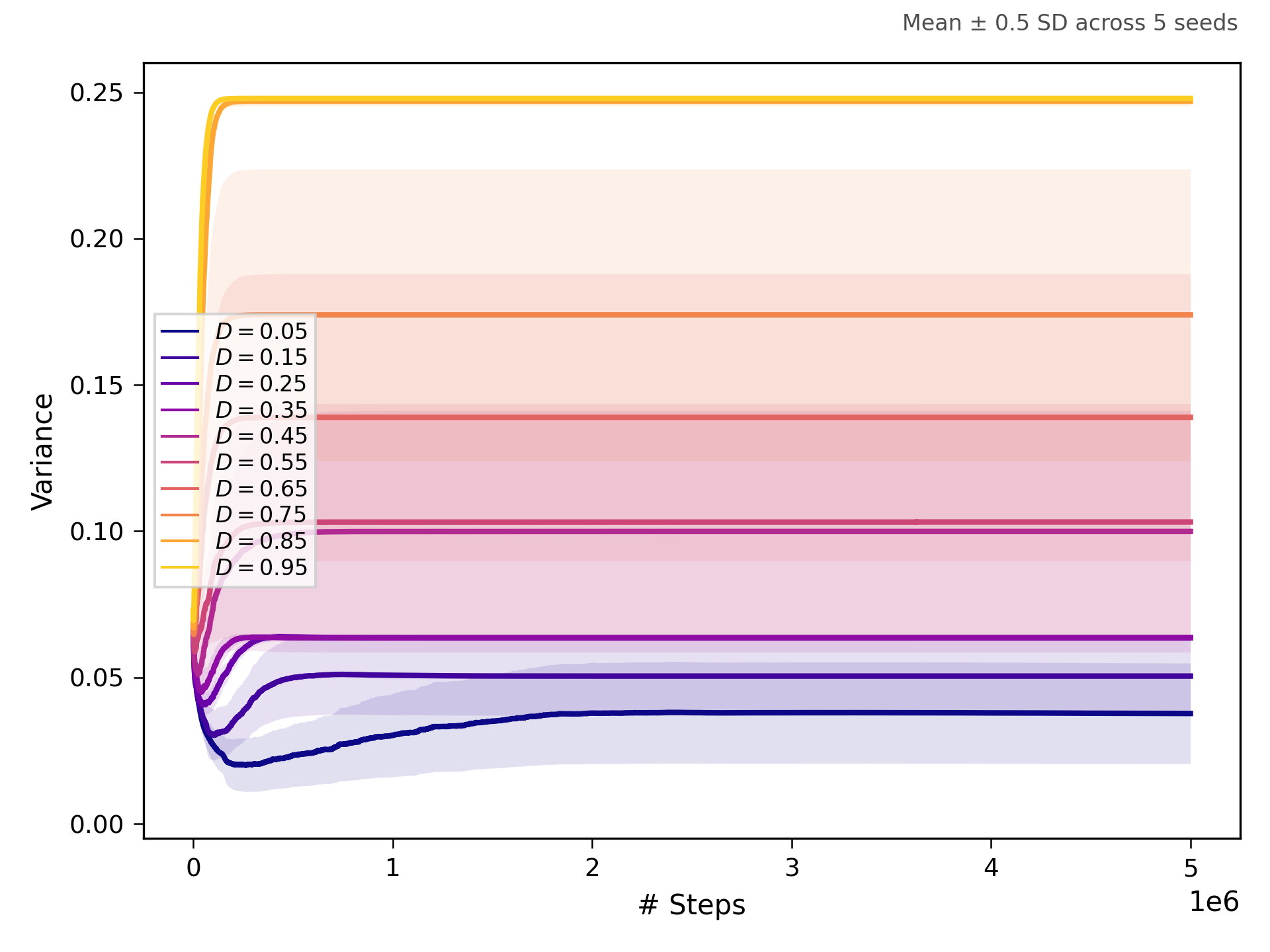}
    \caption{Polarization dynamics under varying diversification parameter $D$ ($E' = 100$, $T = 0.2$, polarized initialization).}
    \label{fig:sweep-D}
  \end{subfigure}
  \caption{Sensitivity analysis of recommendation parameters. (a) Increasing $E'$ suppresses polarization under the tolerance-aware rule. (b) Small $D$ reduces polarization; intermediate $D$ yields intermediate variance, and large $D$ polarizes. Curves: five-seed means; shading: $\pm0.5$ sample SD.}
  \Description{Two line charts of opinion variance against the number of steps. Left: one curve per value of E' from 1 to 100; E' = 1 rises to 0.25, while larger E' stay near 0.06 to 0.10. Right: one curve per value of D from 0.05 to 0.95; small D falls to near 0, intermediate D settles near 0.063, and large D rises to about 0.24.}
  \label{fig:sweep}
\end{figure}

\subsection{Adaptive Diversification}

Although tolerance-aware filtering reduces interactions between agents whose opinion distance exceeds the tolerance threshold, all agents still share the same exposure parameter. Consequently, the degree of personalization remains identical across the population, regardless of an agent's position in the opinion space. To introduce heterogeneous exposure while preserving the personalized interaction mechanism, we allow the exposure parameter to vary adaptively according to an agent's opinion position~\cite{helberger2018exposure,matakos2022maximizing,biega2018equity}.

Without loss of generality, let $x_m$ denote a reference position, which is fixed at $x_m = 0.5$ throughout the simulations. At time $t$, the distance between agent $i$ and the reference position is $d_{m,i}(t) = |x_i(t) - x_m|$.

Each agent is then assigned an individual exposure parameter as
\begin{equation}
  E_i(t) = D^{1/d_{m,i}(t)}, \qquad d_{m,i}(t) > 0,
  \label{eq:adaptive-exposure}
\end{equation}
where $0 < D < 1$ is the diversification parameter. Because $D < 1$, agents located farther from the reference position are assigned larger exposure parameters, resulting in broader interaction opportunities, whereas agents closer to the reference position retain more personalized exposure.

For $d_{m,i}(t) > 0$, the raw interaction probability is $q_{ij}(t) = E_i(t)^{d_{ij}(t)} = D^{d_{ij}(t)/d_{m,i}(t)}$. If $d_{m,i}(t) = 0$, we set $q_{ij}(t) = 1$ for $d_{ij}(t) = 0$ and $q_{ij}(t) = 0$ otherwise: an agent at $x_m$ interacts only with agents at the same position. Combining adaptive diversification with tolerance awareness gives
\begin{equation}
  P_{ij}(t) =
  \begin{cases}
    q_{ij}(t) & d_{ij}(t) \le T,\\
    q_{ij}(t)/E' & d_{ij}(t) > T.
  \end{cases}
  \label{eq:adaptive}
\end{equation}

By calibrating exposure to an agent's position rather than uniformly across the population, Eq.~\eqref{eq:adaptive} preserves personalization as the basis for interaction selection while allowing the degree of personalization to vary adaptively across the opinion space. Larger values of $D$ widen exposure for agents away from $x_m$, whereas smaller values narrow it, most of all for agents near $x_m$. This identical-position restriction at $x_m$ holds for every $D \in (0,1)$.

\begin{figure*}[t]
  \centering
  \begin{subfigure}[t]{0.32\textwidth}
    \centering
    \includegraphics[width=\linewidth]{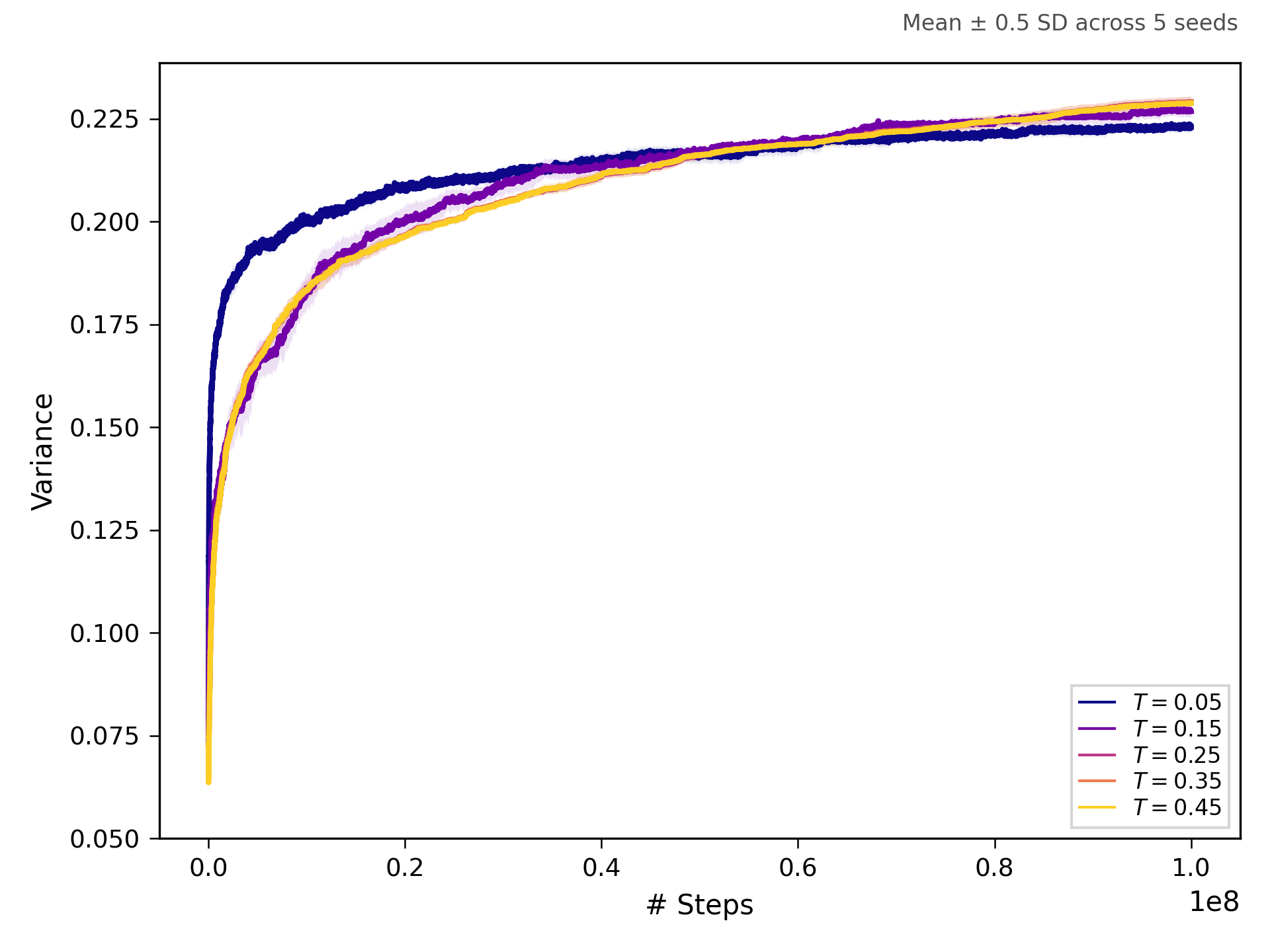}
    \caption{Ultra-personalized recommendations ($E = 0.5^{20}$).}
    \label{fig:designs-personalized}
  \end{subfigure}\hfill
  \begin{subfigure}[t]{0.32\textwidth}
    \centering
    \includegraphics[width=\linewidth]{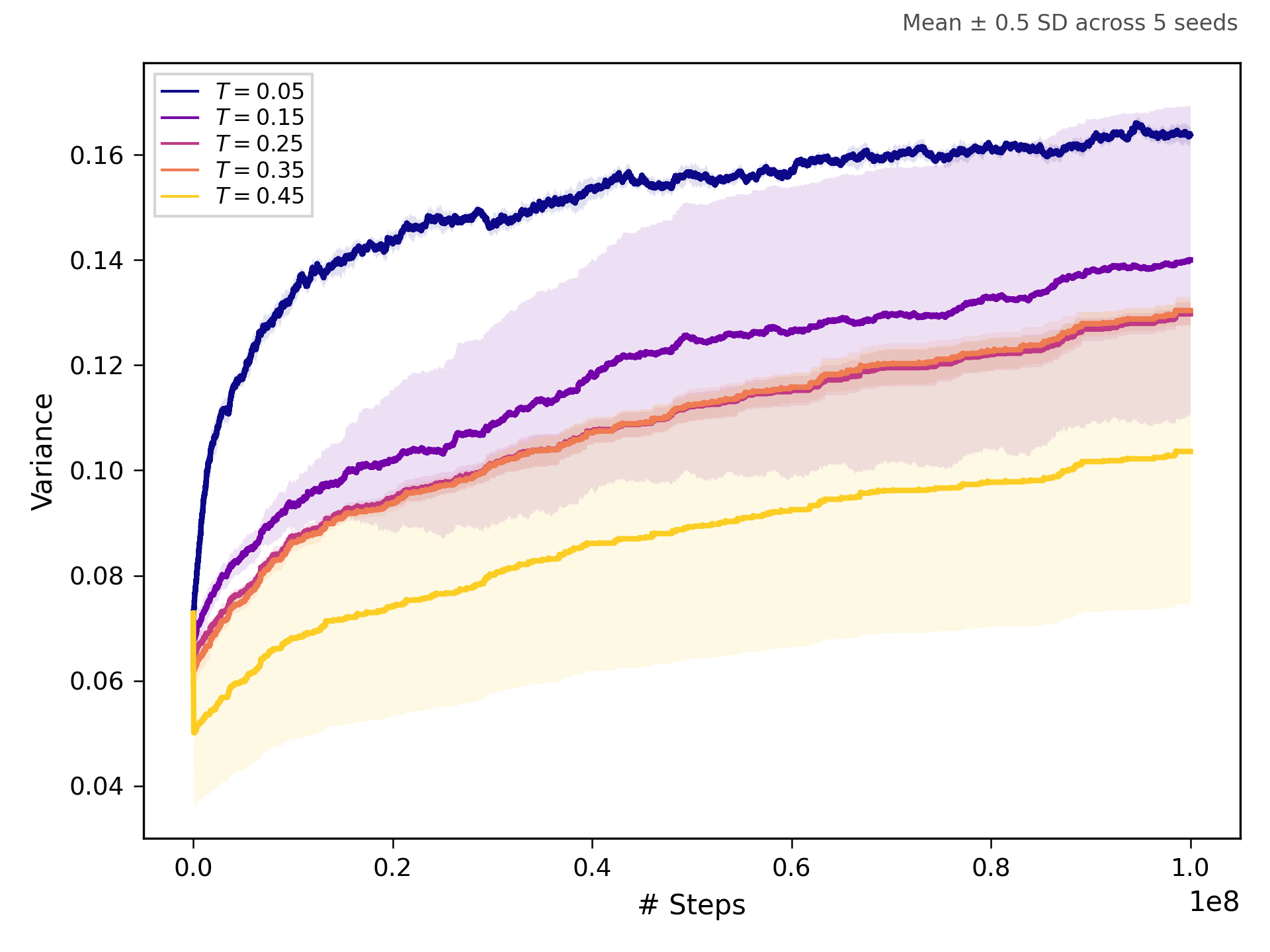}
    \caption{Tolerance-aware recommendations ($E = 0.5^{20}$, $E' = 100$).}
    \label{fig:designs-tolerance}
  \end{subfigure}\hfill
  \begin{subfigure}[t]{0.32\textwidth}
    \centering
    \includegraphics[width=\linewidth]{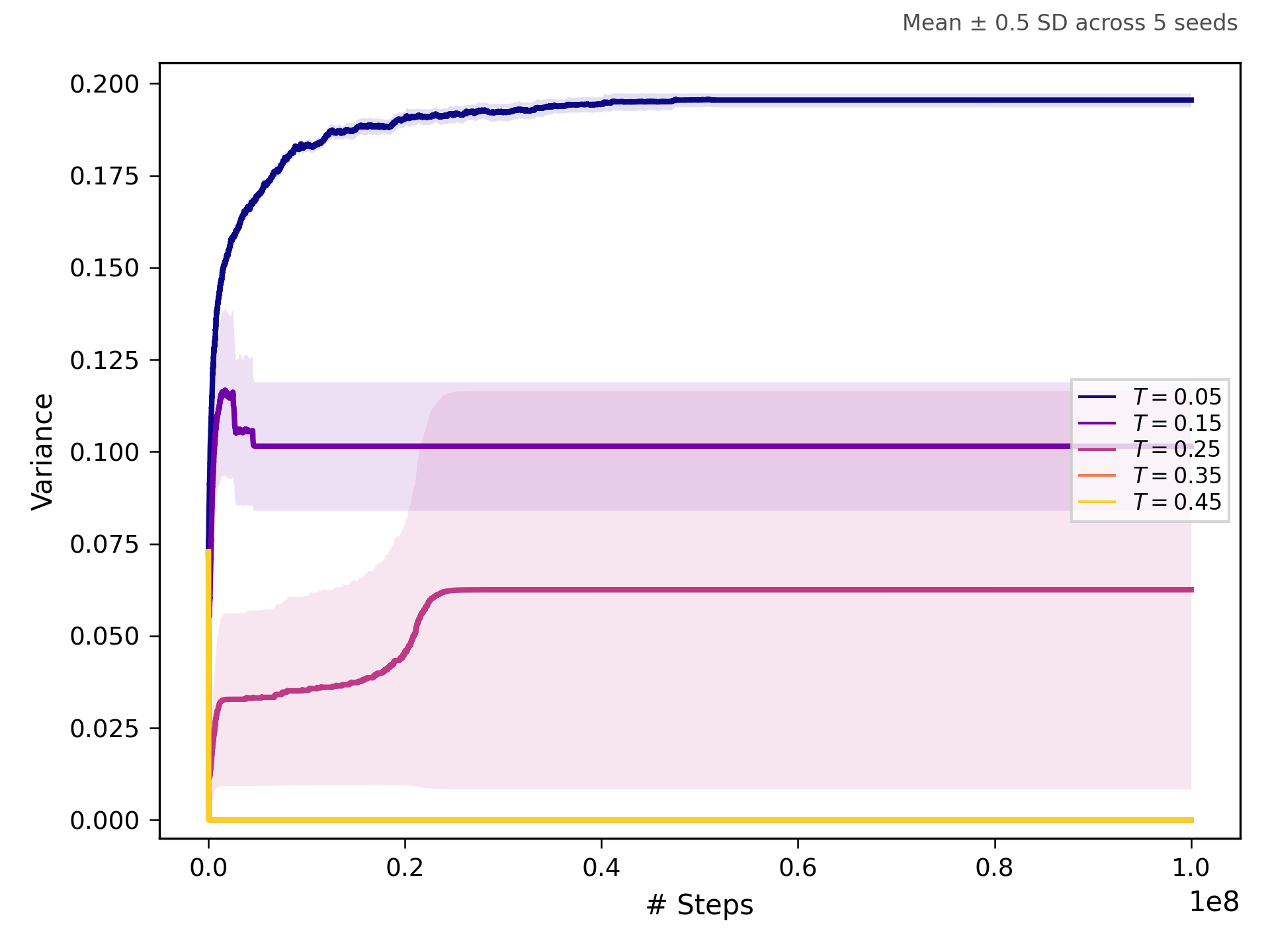}
    \caption{Tolerance-aware and adaptively diversified recommendations ($E' = 100$, $D = 0.15$).}
    \label{fig:designs-adaptive}
  \end{subfigure}
  \caption{Polarization dynamics under different recommender designs, by tolerance $T$, from polarized initializations matched within each seed. Curves: five-seed means; shading: $\pm0.5$ sample SD.}
  \Description{Three line charts of variance against the number of steps, up to 1e8, with one curve per tolerance T from 0.05 to 0.45. Left: all curves rise from about 0.07 to about 0.22. Middle: curves rise more slowly, to between 0.08 and 0.17. Right: T = 0.05 and 0.15 rise to about 0.195 and 0.133, while T of 0.25 and above drop to near 0.}
  \label{fig:designs}
\end{figure*}

\subsection{Modeling Social Pressure}

To incorporate social pressure into the agent-based model, we introduce two complementary mechanisms: the spiral of silence and two fixed extreme influencers. The spiral of silence suggests that individuals who perceive their opinions as inconsistent with the prevailing social climate may become less willing to express or exchange their views. In our model, this mechanism is represented as a reduction in an agent's probability of participating in interactions when its opinion differs from the perceived mainstream position.

For agent $j$, we define a silence factor as
\begin{equation}
  S^{|x_j(t) - x_{*,j}(t)|},
  \label{eq:silence}
\end{equation}
where $S \in [0,1]$ is the silence parameter, with smaller values indicating stronger suppression of opinion expression. Here, $x_j(t)$ denotes the current opinion of agent $j$, and $x_{*,j}(t)$ denotes the perceived mainstream position of agent $j$ at time $t$. The effective probability that agent $i$ interacts with agent $j$ is therefore given by
\begin{equation}
  P_{ij}^{\mathrm{final}}(t) = \min\left(P_{ij}(t), S^{|x_j(t) - x_{*,j}(t)|}\right),
  \label{eq:final}
\end{equation}
where $P_{ij}(t)$ is the interaction probability determined by the recommendation mechanisms described above. This formulation captures the interaction between individual similarity preferences and social pressure: even when two agents have relatively compatible opinions, exposure to agent $j$ may be reduced if $j$ perceives a substantial deviation from the socially salient reference position.

To construct the perceived mainstream position $x_{*,j}(t)$, we introduce two fixed extreme influencers located at the endpoints of the opinion space, with opinions $x_{I_1} = 0$, and $x_{I_2} = 1$.

Unlike regular agents, the influencers remain fixed throughout the simulation and do not update their opinions in response to interactions. Instead, they serve as persistent sources of socially salient opinions. Each influencer broadcasts its opinion with probability $P_{\mathrm{inf}}$, representing its visibility within the simulated social environment. When an influencer becomes accessible through the recommendation process, it can be drawn as a passive partner and its opinion can serve as a social reference for other agents. For agent $j$, the perceived mainstream position $x_{*,j}(t)$ is determined by the nearest visible influencer. Thus, $x_{*,j}(t)$ corresponds to either 0 or 1, depending on which extreme influencer is most salient to agent $j$.

Note the distinction between the fixed center of the opinion space, $x_m = 0.5$, used in the adaptive-diversification mechanism and the perceived mainstream position $x_{*,j}(t)$ generated by the extreme influencers: the former is a fixed geometric reference for calibrating exposure, whereas the latter represents a socially perceived reference that can vary across agents and over time.

Across our simulations, the two extreme influencers are fixed at positions 0 and 1, and their broadcast probability is fixed at $P_{\mathrm{inf}} = 0.15$ to examine opinion dynamics under exposure to extreme social references. Combined with tolerance-aware and adaptive-diversification mechanisms, these components provide an agent-based framework for studying how recommendation-based exposure and socially perceived norms jointly shape collective opinion dynamics.

\begin{figure}[t]
  \centering
  \begin{subfigure}[t]{0.40\linewidth}
    \centering
    \includegraphics[width=\linewidth]{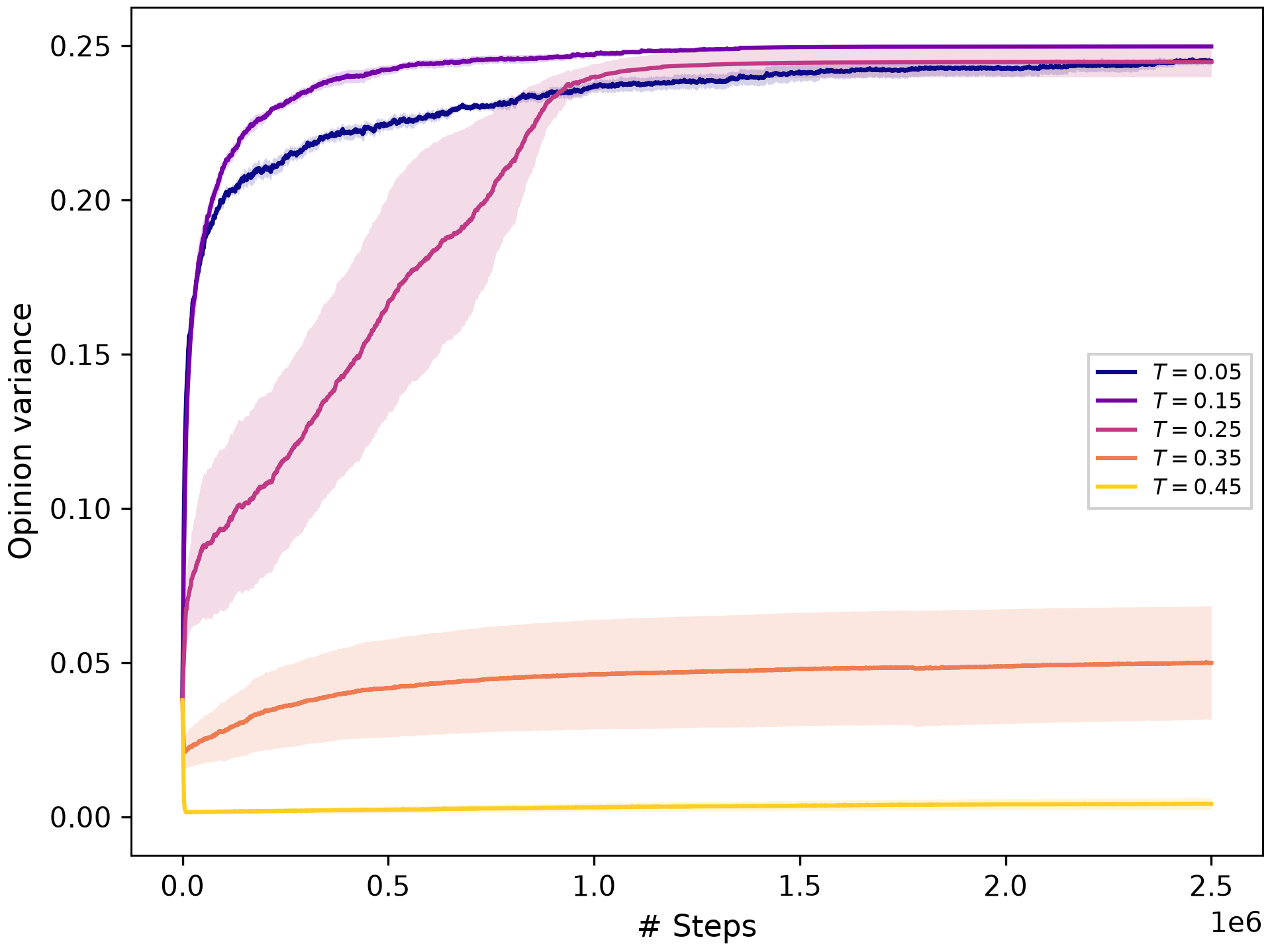}
    \caption{No influencer or spiral effects.}
    \label{fig:pressure-baseline}
  \end{subfigure}\hspace{0.035\linewidth}
  \begin{subfigure}[t]{0.40\linewidth}
    \centering
    \includegraphics[width=\linewidth]{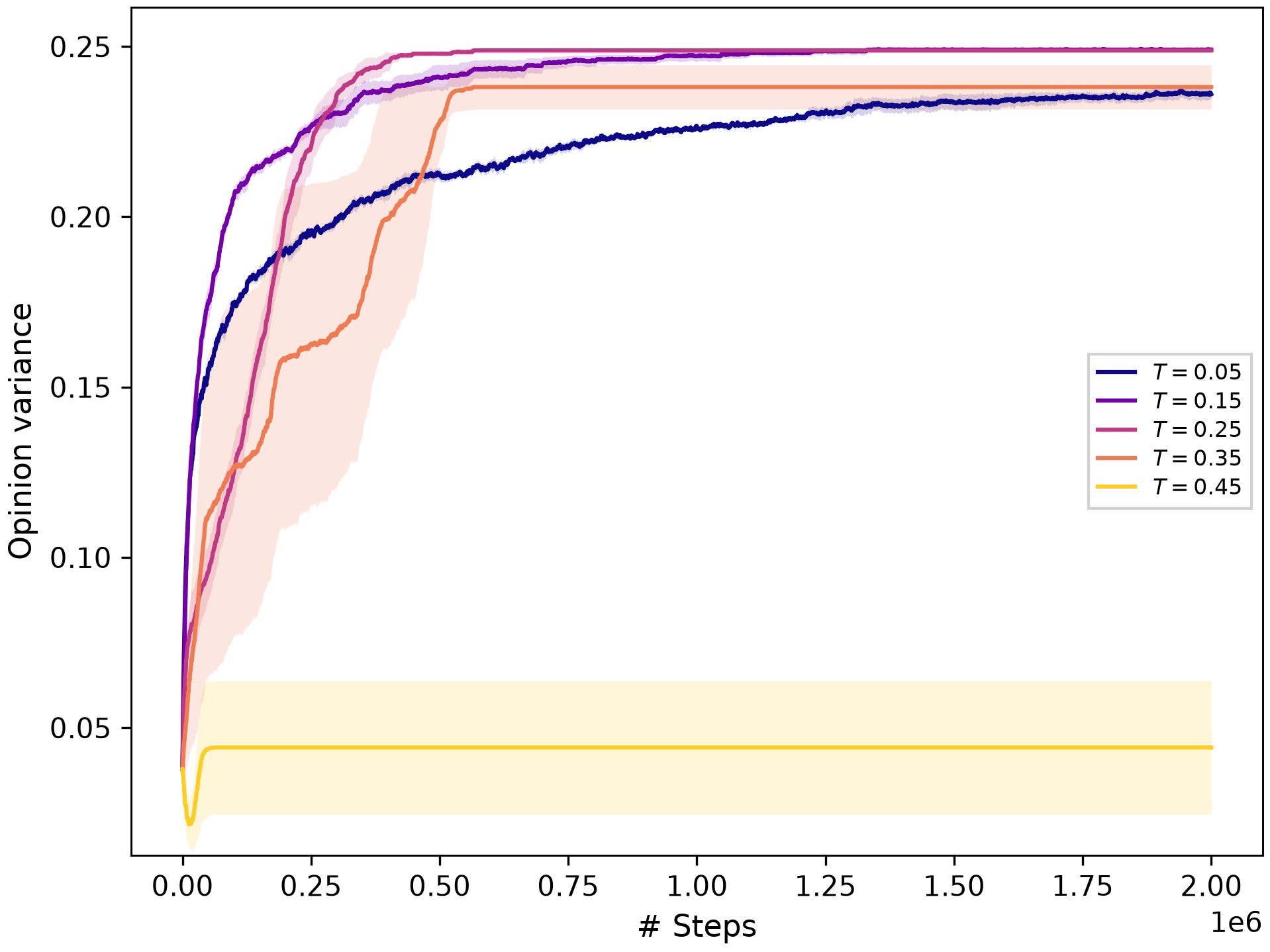}
    \caption{With two extreme influencers ($P_{\mathrm{inf}} = 0.15$).}
    \label{fig:pressure-influencers}
  \end{subfigure}

  \medskip
  \begin{subfigure}[t]{0.40\linewidth}
    \centering
    \includegraphics[width=\linewidth]{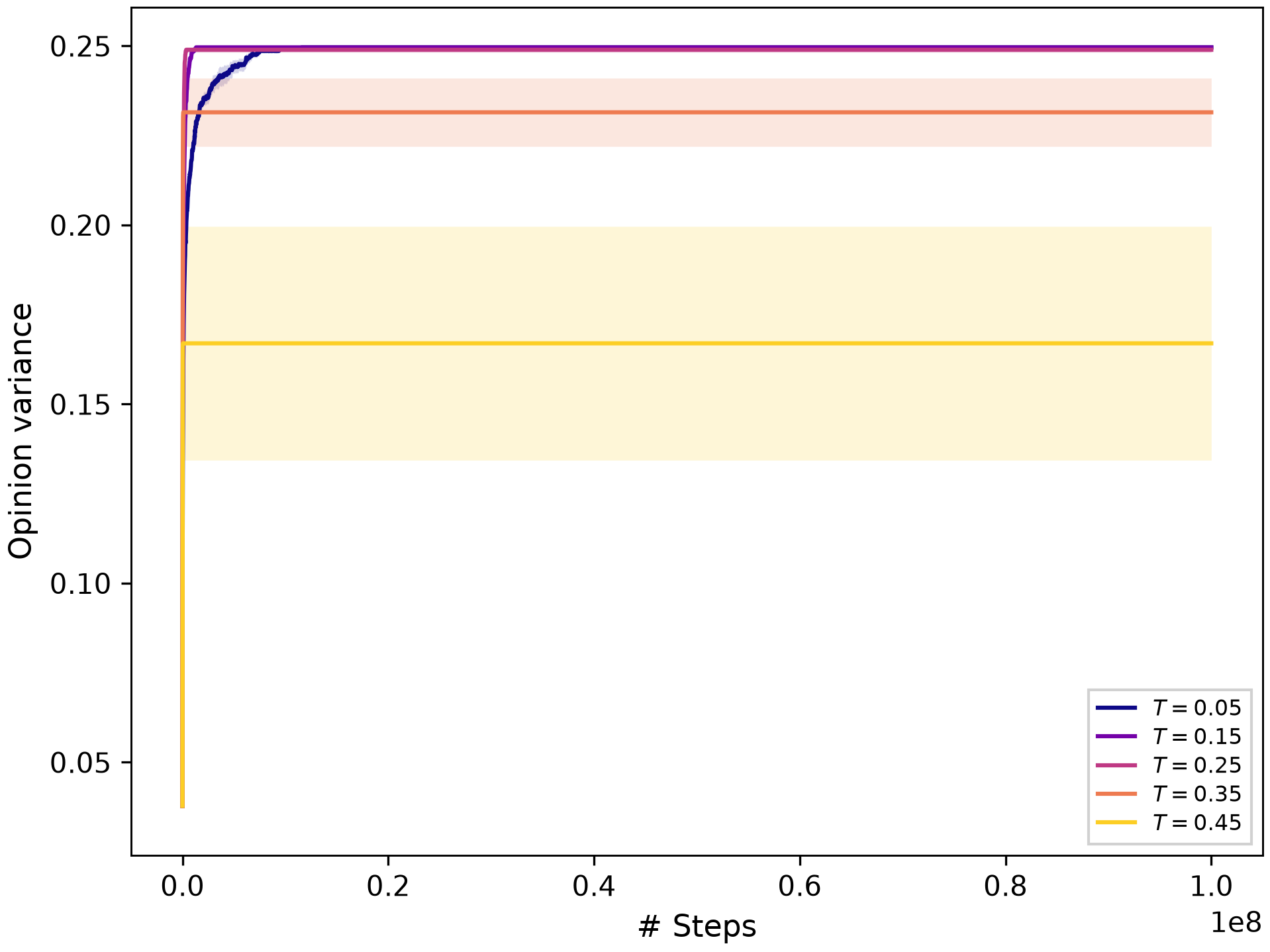}
    \caption{With both influencers and spiral of silence ($S = 0.01$).}
    \label{fig:pressure-silence}
  \end{subfigure}\hspace{0.035\linewidth}
  \begin{subfigure}[t]{0.40\linewidth}
    \centering
    \includegraphics[width=\linewidth]{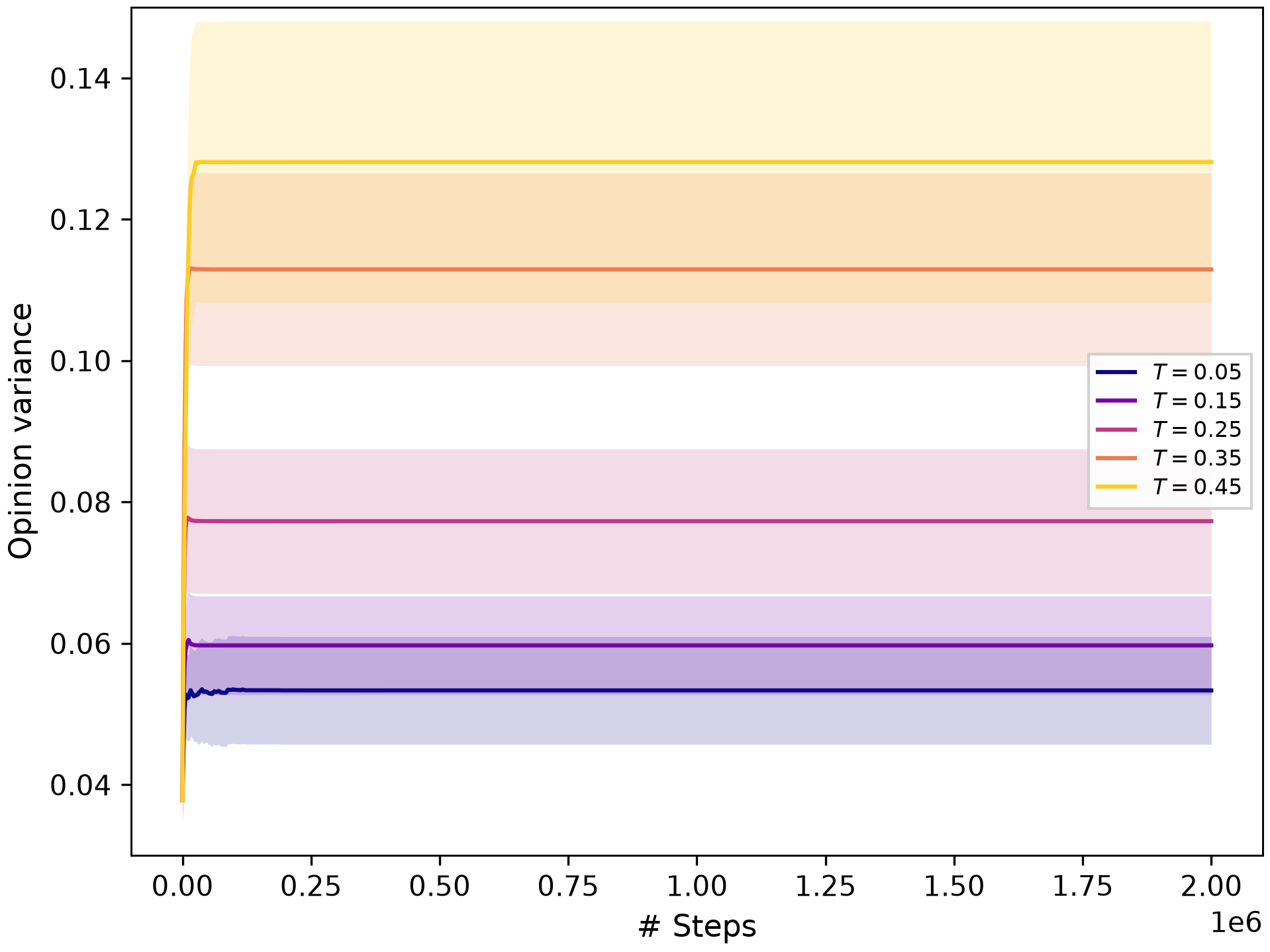}
    \caption{With both effects and proposed recommendation design ($E' = 100$, $D = 0.5$).}
    \label{fig:pressure-proposed}
  \end{subfigure}
  \caption{Combined effects of extreme influencers and the spiral of silence. Fixed-exposure panels (a)--(c) use $E = 0.5^{10}$; panel (d) uses adaptive exposure instead of fixed $E$. Panels (b)--(d) use $P_{\mathrm{inf}} = 0.15$, and panels (c)--(d) use $S = 0.01$. Elevated polarization is observed in Figs.~\ref{fig:pressure-influencers} and~\ref{fig:pressure-silence}, but is mitigated relative to Fig.~\ref{fig:pressure-silence} in Fig.~\ref{fig:pressure-proposed}, demonstrating the robustness of the proposed framework under adversarial social conditions. Curves: means; shading: between-seed variability.}
  \Description{Four line charts of variance against the number of steps, one curve per tolerance T. (a) Without influencers or silence, curves with T up to 0.25 rise to about 0.25, while T of 0.35 or more stays low. (b) With two extreme influencers, every curve except T = 0.45 rises to about 0.25. (c) With influencers and the spiral of silence, all curves jump to between 0.21 and 0.25 almost at once. (d) With the proposed design, the curves settle between 0.04 and 0.14.}
  \label{fig:pressure}
\end{figure}

\subsection{Parameter Settings}

Unless stated otherwise, simulations have $N = 100$ agents and responsiveness $R = 0.25$, $T$ ranges from 0.05 to 0.95, and Figs.~3--5 aggregate multiple seeds; captions give the remaining parameters. Appendix~\ref{app:settings} gives the run lengths, initializations and seeds.

\section{Results}
\label{sec:results}

Four main findings emerge from our simulation experiments: (i) the failure of ultra-personalized recommendations under polarized initial conditions; (ii) the slowing effect of tolerance-aware filtering; (iii) the reversal achieved by adaptive diversification; and (iv) the robustness of the combined framework against spiral-of-silence dynamics.

\subsection{Limit of Personalization}
\label{sec:res-personalization}

We begin by assessing baseline performance under ultra-person\-alized recommendations (exposure parameter $E = 0.5^{20}$). Previous work shows that under a unimodal Gaussian ideological distribution, highly personalized recommendations ($E = 0.5^{20}$) can slow polarization by reducing repulsive cross-cutting interactions~\cite{axelrod2021preventing,sikder2020minimalistic,donkers2021dual,peralta2021effect,blex2022positive}. Whether this protective effect persists when the population is already polarized at initialization remains unclear. Our analysis therefore centers on polarized initial configurations---where opinion variance is already elevated (bimodal distribution, mean initial variance approximately 0.073) and opposing camps are entrenched (see Fig.~\ref{fig:init-bimodal} for an example).

Under these conditions, ultra-personalized recommendations ($E = 0.5^{20}$) fail to reverse polarization and may instead reinforce existing boundaries. As shown in Fig.~\ref{fig:designs-personalized}, for low-to-intermediate tolerance ($T \le 0.45$), polarization increases over time. Although repulsive interactions occur infrequently under low exposure, the large initial distances between opposing clusters mean that when repulsion does occur, the positional shift is substantial, driving agents toward the extremes. Attractive interactions, by contrast, occur primarily within each cluster and tend to maintain the initial polarized configuration. Personalization-based mitigation thus exhibits a fundamental limitation: its efficacy depends critically on the initial ideological distribution.

\subsection{Tolerance-Aware Recommendation Slows Polarization}
\label{sec:res-tolerance}

Can incorporating tolerance into interaction selection overcome the limitations of personalization alone? Figure~\ref{fig:sweep-Eprime} shows that introducing tolerance awareness substantially slows the growth of polarization from a unimodal initial distribution. Even a modest level of tolerance awareness markedly suppresses the increase in opinion variance relative to the baseline without tolerance awareness, leading to consistently lower polarization throughout the simulation.

Evaluating the mechanism under an initially polarized population with highly personalized exposure reveals a similar benefit. As shown in Fig.~\ref{fig:designs-tolerance}, tolerance-aware recommendation consistently reduces polarization compared with the corresponding baseline (Fig.~\ref{fig:designs-personalized}) across a broad range of tolerance thresholds. Because this improvement is observed for both initially homogeneous and already polarized populations, this relative benefit is robust across the tested initial configurations.

Despite these improvements, mean trajectories remain polarized at the simulated horizon. Tolerance-aware recommendation slows the development of polarization and lowers its overall level, but it does not consistently restore polarized populations to a low-polarization state across seeds. Motivating the next stage of our design, this limitation points to the need for the adaptive diversification mechanism introduced in the following section.

\subsection{Adaptive Diversification Reverses Polarization}
\label{sec:res-adaptive}

While tolerance-aware recommendation substantially slows polarization, it remains insufficient to consistently restore already polarized populations across seeds; we therefore ask whether adaptive diversification can reverse polarization once fragmented opinion distributions have formed.

Figure~\ref{fig:designs-adaptive} demonstrates that adaptive diversification fundamentally changes the long-term dynamics. From initial configurations matched within each seed, the combined mechanism reduces final mean polarization below its initial level for $T \ge 0.25$. Consensus occurs in 5/5 runs at $T = 0.35, 0.45$, versus 3/5 at $T = 0.25$. For $T = 0.05$ and 0.15, however, the mean variance rises from approximately 0.073 to about 0.195 and 0.102, respectively. These are respectively above and below the tolerance-aware-only means (about 0.164 and 0.140). Unlike tolerance-aware recommendation alone, which generally lowers the level of polarization, adaptive diversification enables polarized populations to recover from fragmented states under suitable conditions.

To clarify this transition, Fig.~\ref{fig:sweep-D} traces the opinion dynamics across different diversification strengths. Reversal occurs at small $D$ (0.05 and 0.15 in Fig.~\ref{fig:sweep-D}); for $D$ from 0.25 to 0.55 the mean variance settles between approximately 0.064 and 0.103, and for $D = 0.85$ and 0.95 the population polarizes, with final mean variance approximately 0.247--0.248. Broader exposure from a larger $D$ therefore does not necessarily yield lower polarization.

Collectively, these findings demonstrate that adaptive diversification not only mitigates polarization but can also reverse established polarization, with consistent consensus across seeds under suitable tolerance levels.

\subsection{Robustness to the Spiral-of-Silence Mechanism}
\label{sec:res-silence}

To test whether the proposed framework remains effective when opinion dynamics are further influenced by social pressure, all simulations in this section start from the Gaussian initial distribution and, unless otherwise specified, include two extreme influencers located at the opinion boundaries (0 and 1). Each influencer broadcasts with probability $P_{\mathrm{inf}} = 0.15$, and the spiral-of-silence parameter is set to $S = 0.01$, corresponding to a strong social-pressure condition.

Baseline dynamics under personalized interaction alone appear in Fig.~\ref{fig:pressure-baseline}. Adding extreme influencers accelerates polarization (Fig.~\ref{fig:pressure-influencers}), and the spiral of silence amplifies it further for some $T$ (Fig.~\ref{fig:pressure-silence}).

Deploying the proposed framework under the same social-pressure setting substantially counteracts this amplification. As shown in Fig.~\ref{fig:pressure-proposed}, combining tolerance-aware recommendation with adaptive diversification reduces mean polarization across the tested tolerance levels relative to the corresponding baseline with the spiral-of-silence mechanism. Notably, this reduction remains evident despite the presence of both influencers and strong social pressure.

Overall, these experiments confirm that the proposed framework remains robust when polarization is reinforced by social pressure. Although the spiral-of-silence mechanism can substantially increase polarization, the combined interaction mechanisms continue to mitigate its effects and maintain substantially lower mean polarization than the matched personalized baseline.

\section{Testing the Mechanism with LLM-Powered Agents}
\label{sec:llm}

We examine whether the recommendation effects in Section~\ref{sec:results} persist when LLM-generated opinion changes replace the ARM update rule. The agents are explicitly told that nearby views can persuade them and distant views can push them away (P1); the LLM determines the direction and magnitude of each response. Feed selection uses known agent positions and a single estimated population threshold, adapting the ARM's shared-tolerance and known-state assumptions (P2 and P3). This comparison tests the policy mechanisms under a shared behavioral premise, extending work on LLM-based feed and opinion simulations~\cite{tornberg2023simulating,chuang2024simulating}.

\begin{figure*}[t]
\centering
\includegraphics[width=\textwidth]{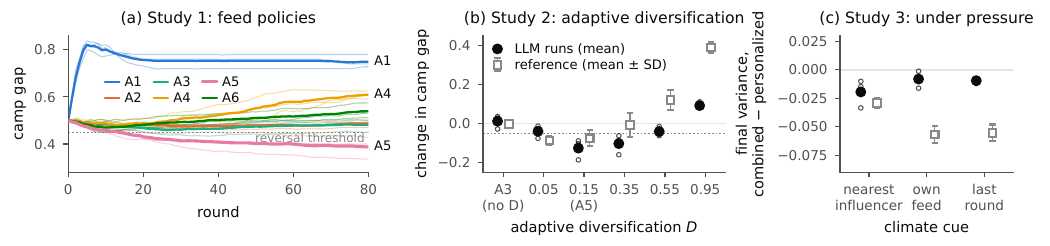}
\caption{Results with LLM agents, three runs per feed. (a) Camp gap in Study~1 (A1--A6 defined in Section~\ref{sec:llm}); bold: mean of three runs, thin: single runs. (b) Change in camp gap after 160 rounds of Study~2 by $D$; black: LLM runs (dot: mean, circles: single runs), gray: rule-based reference (mean $\pm$ SD, 10 seeds), dotted: reversal threshold. (c) Study~3: final variance, combined design minus personalized feeds, under influencers and silence, by climate cue.}
\Description{Three panels. (a) Line chart of the camp gap over 80 rounds for six feeds, the mean of three runs in bold and each run thin: the random feed A1 rises from 0.50 to about 0.75, adaptive diversification alone A4 to about 0.61, the tolerance-aware random feed A6 ends near 0.54, ultra-personalized A2 and tolerance-aware A3 stay near 0.48, and the combined design A5 falls steadily to about 0.39, the only mean curve below the reversal threshold of 0.45. (b) Change in camp gap after 160 rounds by D, mean of three LLM runs against the rule-based reference: A3 +0.01 against 0.00; D = 0.05, -0.04 against -0.09; D = 0.15, -0.13 against -0.08; D = 0.35, -0.10 against -0.01; D = 0.55, -0.04 against +0.12; D = 0.95, +0.09 against +0.39. (c) Final variance of the combined design minus personalized feeds, mean of three LLM runs against the reference, by climate cue: nearest influencer -0.020 against -0.029, own feed -0.008 against -0.057, last round -0.010 against -0.055.}
\label{fig:llm}
\end{figure*}

\subsection{Experimental Setup}

The platform contains 100 persona-based agents discussing a fictional state's retirement contribution rate, ranging from 4\% to 24\% of wages. Rates are mapped to $[0,1]$ by $x=(p-4)/20$. Agents post their views and read up to three sampled posts per round, with display probabilities determined by the recommendation policy. A 20-round pilot estimates a common attraction--repulsion threshold, $\hat T=0.175$, used in all runs. Preliminary trials revealed a directional bias: agents moved farther toward nearby posts favoring higher contribution rates than toward equally distant posts favoring lower rates. This asymmetry could produce upward drift even under balanced exposure. We used randomized mirrored reading to control this asymmetry; the procedure and eligibility restrictions are described in Appendix~\ref{app:llm-platform}. Implementation, calibration, prompts, and the workflow are in Appendix~\ref{app:llm}.

Study~1 compares six feeds from a polarized initialization over 80 rounds: random (A1), ultra-personalized (A2), tolerance-aware (A3), adaptive diversification alone (A4), the combined design (A5), and tolerance-aware random (A6). Study~2 varies diversification strength over 160 rounds. Study~3 examines influencers and silence from a unimodal initialization over 30 rounds, using three opinion-climate cues. Each arm has three LLM runs with different seeds and 10 rule-based reference runs using the same feed rules and initialization distributions, with independently drawn starts. We report opinion variance and define reversal as a reduction of at least 0.05 in the gap between the two initial camps' mean opinions. The comparisons test persistence under personalization (C1), mitigation with tolerance filtering (C2), reversal with the combined design (C3), failure of exposure alone (C4), and mitigation under social pressure (C5).

\subsection{Results and Scope}

In Study~1, personalized and tolerance-aware feeds barely changed the camp gap ($-0.013$ and $-0.019$ on average over three runs), although tolerance awareness kept final variance slightly lower in every run (0.061 versus 0.064). Only the combined design met the reversal criterion in all three runs, reducing the gap by 0.113 on average (0.070--0.166) and ending at variance 0.041, compared with 0.105 under adaptive diversification alone (Fig.~\ref{fig:llm}a). Random exposure widened the gap in every run ($+0.249$). As in Section~\ref{sec:results}, tolerance awareness alone did not reliably reverse polarization, whereas combining it with adaptive diversification did.

The parameter sweep broadly follows Fig.~\ref{fig:sweep-D} (Fig.~\ref{fig:llm}b). At $D=0.15$ and 0.35, the gap reductions persisted through 160 rounds in all three runs ($-0.127$ and $-0.103$ on average), and excessive diversification ($D=0.95$) widened the gap in every run ($+0.091$). At $D=0.05$, an early narrowing partly faded ($-0.040$ on average; one run stayed reversed), whereas the rule-based reference reversed in 9 of 10 seeds. Under influencers and silence, the combined design lowered final variance under all three cues in every run, by 0.008--0.020 on average, versus 0.029--0.057 in the reference (Fig.~\ref{fig:llm}c). These scenarios test mitigation from a unimodal population; adding influencers did not increase variance over the 30-round horizon.

The results support a conditional policy comparison: the combined design retains its advantage with LLM-generated updates, while the response to weak diversification and the size of the social-pressure benefit differ from the rule-based model. The supplied premise, symmetrization, and outcome-based gates limit independent validation; three runs per arm show consistent directions but only roughly estimate outcome variability. Full arm-level outcomes are in Appendix~\ref{app:results}.

\section{Discussion}

\subsection{Mechanism-Level Insights}

This study highlights a central tension in recommender-system design: personalization can limit destabilizing disagreement while also restricting opportunities for already divided groups to move closer. Algorithms that maximize engagement through preference alignment—such as those deployed by YouTube, TikTok, X and other content platforms—are increasingly implicated in the amplification of homophily and the erosion of cross-cutting exposure \cite{sunstein_republic_2018,cinelli_echo_2021,huszar_algorithmic_2022,santos_link_2021}. While personalization satisfies immediate cognitive comfort and increases retention, it inadvertently fosters epistemic fragmentation \cite{flayelle_taxonomy_2023,cinelli_echo_2021}. Our results suggest that the solution is not to reject personalization but to modulate it through mechanisms responsive to user-specific thresholds and system-level ideological dynamics. 

Central to our argument is the recognition that ideological tolerance is a latent but critical user attribute that governs the downstream impact of exposure diversity. Existing recommender systems tend to model users as monolithic in their receptivity to divergent views \cite{kozyreva_citizens_2020}. However, both psychological theory and our simulations suggest that this is a flawed assumption: in low-tolerance environments, even modest increases in ideological diversity can trigger repulsion, identity threat, or accelerated radicalization \cite{axelrod_dissemination_1997,noorazar_classical_2020,bail_exposure_2018,peralta_multidimensional_2024}. Our proposed tolerance-aware mechanism addresses this by adjusting exposure probabilities based on ideological distance relative to tolerance thresholds. 
This approach reduces the likelihood of repulsive encounters while preserving in-threshold heterogeneity—creating informational environments conducive to bounded pluralism rather than polarized bifurcation. 

Yet, we find that mitigating polarization is not equivalent to reversing it. In systems already characterized by entrenched fragmentation, tolerance-aware filtering merely slows polarization’s progression; it does not alter its directionality \cite{bail_exposure_2018}. To address this, we introduce adaptive diversification, a structurally asymmetric mechanism that increases recommendation diversity as a function of ideological extremity. This strategy broadens exposure for agents farther from the reference position while retaining stronger personalization near it. Our simulations reveal a nonlinear relationship between diversification and polarization: appropriately calibrated diversification can promote convergence or reduce opinion dispersion, whereas excessive diversification can intensify polarization. This underscores a central finding: polarization dynamics are sensitive to second-order design parameters, not merely the presence or absence of diversity. 
 
 These mechanisms also highlight the necessity of designing for heterogeneity, not just averages. Modern platforms often rely on aggregate-level performance metrics (e.g., click-through rate, dwell time), obscuring the asymmetric effects that algorithmic exposures have on marginal versus moderate users. By simulating agents with distinct ideological positions and tolerance thresholds, our framework recovers a more ecologically valid understanding of population dynamics. In particular, the disproportionate influence of peripheral agents—who, despite being few in number, shape system-level extremity through high-amplitude repulsion—is an emergent property missed by homogeneous models. 
 
 We further enrich our model by formalizing the spiral of silence \cite{noelle-neumann_theory_1991}, a theory long associated with mass communication but rarely implemented in algorithmic systems modeling. Our approach captures how perceived mainstream positions—whether formed through influencers, media signals, or feedback loops—can suppress dissent and reshape participation landscapes. This pattern has been especially evident during election periods in polarized media environments, where expressive suppression intensifies social conflict \cite{gearhart_spiral_2025}. The silencing mechanism attenuates interaction probabilities based not only on ideological distance but also on positional salience. Marginalized views are not only algorithmically downranked but also self-censored, reinforcing the visibility of dominant positions and narrowing the expressive bandwidth of public discourse \cite{medlar_nobody_2022,gillani_me_2018}. Such silencing is shaped not only by immediate social cues, but also by shared representational fields that cognitively anchor what is perceived as mainstream.
 
Importantly, our results show that even under such adversarial social conditions, the proposed tolerance-aware and adaptive diversification framework remains robust. When coupled with spiral-of-silence effects and extreme influencer presence, the joint mechanism stabilizes ideological trajectories and curtails the upward spiral of polarization. Each mechanism addresses a distinct structural vulnerability: tolerance-aware filtering stabilizes short-range interactions; adaptive diversification reintegrates ideological outliers; spiral-of-silence modeling accounts for expressive asymmetry.

\subsection{Modeling Limitations and Future Work}

Despite these strengths, the framework has several limitations that open avenues for future work. First, we assume that the initially polarized population still includes individuals with intermediate opinions, who may provide bridges between the two camps. Whether the proposed mechanisms remain effective when such intermediaries are scarce or absent requires further investigation. Second, our implementation assumes static tolerance parameters and symmetric responsiveness, while real-world users evolve over time. Modeling tolerance as an endogenous variable—shaped by prior exposures, emotional valence, or network embedding—could yield richer dynamics and allow for the investigation of resilience thresholds or tipping points. Third, our model operates in a unidimensional ideological space, abstracting from issue-specific polarization or cognitive dissonance between overlapping identity axes. Extending the model to multidimensional opinion landscapes \cite{peralta_effect_2021} would enhance its realism and applicability to real-world sociopolitical systems. Empirical work has shown that exposure to divergent views across multiple identity dimensions can backfire via affective polarization \cite{bail_exposure_2018}, while even homogeneous interactions can produce bipolarization under certain structural conditions \cite{mas_differentiation_2013}.

Although the rule-based model assumes fixed tolerance and responsiveness parameters, the LLM experiments relax these behavioral restrictions by introducing agents with diverse backgrounds and levels of issue involvement, whose tolerance and opinion changes are generated in response to their personas, current views, and encountered content. They therefore provide a complementary test beyond the prescribed numerical update rule. Across three runs, they reproduced the central ordering of the policies: only the combined design reversed polarization in every run, although its benefit under social pressure was smaller than in the rule-based model. For future work, independent evaluation across prompts, models, and issues, together with human studies, is needed to establish broader behavioral applicability.

A particularly promising direction lies in the incorporation of multiplex mainstream perception. Currently, each agent’s perceived mainstream position is approximated via proximity to a fixed influencer, which may not reflect the fragmented and siloed realities of digital publics \cite{bakshy_exposure_2015}. The lack of algorithmic transparency in mainstream modeling \cite{mittelstadt_explaining_2019} further compounds the difficulty of evaluating salience. Realistically, users are embedded in distinct interpretive communities, each with its own semantic and normative scaffolding. Modeling network-localized mainstreams or algorithmically emergent salience layers (e.g., trending topics, comment sentiment distributions) would allow us to assess the resilience of consensus under multi-polar dominance conditions. 

While validating these mechanisms in real-world platforms remains challenging due to limited API access and user data, prior field experiments on ideological exposure \cite{guess_digital_2020} offer promising templates. Hybrid agent-user simulations and natural experiments—such as ranking perturbations—could bridge theory and deployment. 

We also encourage future research to conduct field deployments and controlled A/B testing of tolerance-aware and diversity-sensitive algorithms across real platform contexts. Hybrid experimental frameworks—such as human-in-the-loop simulations and agent-based system emulators—could bridge the gap between theoretical modeling and real-world behavioral dynamics. Such validation is essential for establishing construct validity, calibrating real-world design thresholds, and informing the normative trajectories of algorithmic governance.

\subsection{Policy Implications}
Our findings motivate empirical testing of exposure policies calibrated to users' receptivity to disagreement. First, platforms shall incorporate tolerance-aware filtering to prevent counterproductive exposure beyond users’ psychological thresholds \cite{nelimarkka_social_2018}. By dynamically adjusting exposure probabilities based on inferred receptivity, platforms can preserve engagement without triggering repulsive polarization.

Second, adaptive diversification mechanisms should be targeted toward users at extremes, rather than applied uniformly. Algorithmic diversity must be context-sensitive, not absolute \cite{bail_exposure_2018,peralta_effect_2021}. Centrists benefit from relevance and stability; extremists require calibrated nudges toward epistemic moderation. 

Third, addressing the spiral of silence requires that platforms go beyond content-based optimization. Visibility algorithms should account for social pressure dynamics—ensuring that dissenting but reasonable voices are not structurally silenced. Recommender systems shall integrate perceived mainstream modeling and adjust amplification accordingly. 

Collectively, these recommendations suggest that algorithmic neutrality is insufficient. Recommender systems shall be proactively optimized for epistemic resilience and ideological inclusion \cite{binns_fairness_2018,roux_ethics_2023}. Regulatory frameworks and platform guidelines should adopt exposure diversity not as a fixed quota, but as an adaptive, behaviorally grounded design variable. Platforms can model and report on polarization metrics as part of algorithmic impact assessments. Our model provides a theoretically grounded and empirically testable framework for such evaluations.

\section*{Ethics and Disclosures}

\paragraph{Ethical considerations.} This study involves no human participants or personal data; all personas are synthetic, and the state and policy question are fictional. Because tolerance-aware and adaptive exposure deliberately shape what users see, deployment would require transparency, user consent and evaluation with people.

\paragraph{Code and generative AI\@.} Code, prompts and simulation outputs will be released after review. Beyond the LLM agents of Section~\ref{sec:llm}, generative AI tools helped draft and edit parts of the text and code; the authors take full responsibility for all content.

\bibliographystyle{plainnat}
\bibliography{references-arxiv}

\appendix

\section{Simulation Settings of the Rule-Based Model}
\label{app:settings}

Unless stated otherwise, simulations have $N = 100$ agents, and each curve shows a single run with experiment seed 3121127542. The run seed is drawn from NumPy's default random generator initialized with this experiment seed. One step is one randomly drawn pairwise interaction attempt, including rejected attempts; the horizontal axis gives the number of such steps. Figure~\ref{fig:sweep-Eprime} and Fig.~\ref{fig:pressure-baseline} run for $2.5 \times 10^6$ steps; Fig.~\ref{fig:sweep-D} for $5 \times 10^6$; all panels of Fig.~\ref{fig:designs} for $10^8$; Figs.~\ref{fig:pressure-influencers} and \ref{fig:pressure-proposed} for $2 \times 10^6$; and Fig.~\ref{fig:pressure-silence} for $10^8$. Unimodal initialization uses a Gaussian with mean 0.5 and SD 0.2, restricted to $[0,1]$ by rejection sampling. Polarized initialization draws 50 opinions from $N(0.8, 0.1^2)$ and 50 from $N(0.3, 0.1^2)$, then clips them to $[0,1]$; its sampled variance is approximately 0.074 (Fig.~\ref{fig:init}). This initialization is not symmetric about $x_m = 0.5$. Responsiveness is $R = 0.25$, and $T$ ranges from 0.05 to 0.95. Fixed-exposure simulations use $E = 0.5^{20}$ in Fig.~\ref{fig:designs} and $E = 0.5^{10}$ in Fig.~\ref{fig:sweep-Eprime} and the fixed-exposure panels of Fig.~\ref{fig:pressure}; adaptive exposure replaces fixed $E$ wherever diversification is active. The tolerance-awareness parameter $E'$ ranges from 1 to 100 and $D$ from 0.05 to 0.95. Social-pressure simulations use $P_{\mathrm{inf}} = 0.15$ and, when silence is active, $S = 0.01$.

\section{Details of the LLM-Powered Tests}
\label{app:llm}

\subsection{Agents, Feeds, and Controls}
\label{app:llm-platform}

\begin{figure*}[t]
\centering
\includegraphics[width=\textwidth]{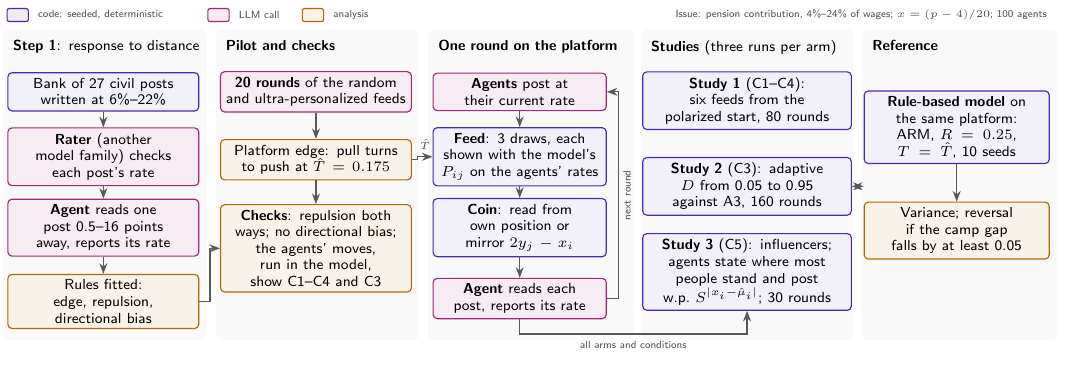}
\caption{Workflow of the conditional LLM tests. Single-post trials (Step~1) and a 20-round pilot measure responses and gate the main runs on compatibility with P1 and C1--C4. Feeds use the model's selection rules; eligible reads use the agent's own or mirror position with equal probability. Each arm has three LLM runs, compared with 10 rule-based reference runs.}
\Description{A flow diagram in five columns. Step 1: a bank of 27 civil posts written at 6 to 22 percent is checked by a rater from another model family; agents read one post at a time at distances from 0.5 to 16 points and report their rates; update rules are fitted. Pilot and checks: 20 rounds of the random and ultra-personalized feeds give the platform edge, T hat = 0.175, and the checks require repulsion in both directions, no directional bias, and that the agents' moves show C1 to C4 and C3 in the model. One round on the platform: agents post at their current rate; a feed makes three draws, each shown with the model's probability; a coin decides whether each post is read from the agent's own or the mirror position; agents read each post and report their rates; then the next round starts. Studies: Study 1 (six feeds, 80 rounds), Study 2 (adaptive D from 0.05 to 0.95 against A3, 160 rounds) and Study 3 (influencers and silence, 30 rounds), three runs each. Reference: the rule-based model on the same platform, with R = 0.25 and T = T hat over 10 seeds; outcomes are the variance and a reversal if the camp gap falls by at least 0.05.}
\label{fig:llm_workflow}
\end{figure*}

\paragraph{Agents and issue.}
One hundred persona-based agents~\cite{argyle2023out}, each running on DeepSeek V4.1 Flash (API identifier \texttt{deepseek-flash}) with reasoning disabled and temperature 0.7, populate the simulated platform. Modelled as residents of Nova, a fictional U.S. state, agents have an age, occupation, place of residence and short background, without party identification. Discourse centers on one issue from our planned human experiment: the share of wages that workers must contribute to a public retirement fund, which the legislature can set between 4\% and 24\%. Mapping a rate $p$ to $x = (p-4)/20$ equates one percentage point to 0.05. Prompts frame the trade-off symmetrically, as larger pensions later at the cost of lower take-home pay now, or the reverse. View-reporting prompts state receptivity to nearby arguments and possible repulsion from distant ones, without specifying a numerical threshold or change size. The exact wording is given in Appendix~\ref{app:prompts}.

\paragraph{Feeds.}
In each round, every agent posts at its current rate; a previous post is reused while that rate is unchanged after rounding to 0.1 percentage point. Three candidate draws per reader each select an available post at random and display it with the recommender's probability $P_{ij}$ (Section~\ref{sec:model}), evaluated at the reader's rate at the start of the round and the post's rate. Agents read displayed posts sequentially and report a new rate after each. Feed selection, random draws and measurement are handled in code (Fig.~\ref{fig:llm_workflow}). The platform regularizes adaptive exposure as $E_i = D^{1/\max(d_{m,i},0.02)}$. For each arm, 10 rule-based reference runs use the same feed rules, initialization distributions and number of rounds, with independently drawn starts and the attraction--repulsion update~\cite{axelrod2021preventing} at $R = 0.25$ and $T = \hat T$.

\paragraph{Response to distance.}
Before the platform runs, agents read single posts from a bank of 27 civil posts written at 6--22\%, checked by a rater from another model family, Gemini 3.8 Flash (\texttt{google/gemini-3.8-flash}). Readers held rates on the same grid or 0.5--1.5 points from the post, and with 90 no-post controls this gave 1,224 trials. Of six update rules fitted to the grid trials, a rule with fixed step sizes fit best (Appendix~\ref{app:gate}). Agents moved about 0.7 points toward posts within 5 points ($T = 0.25$) and about 0.9 points away from more distant posts, in both directions, and kept their rate when shown no post. Close-range trials revealed a directional asymmetry absent from ARM. Among readers at least 2 points from either end of the scale, an agent 0.5 points below a post adopted its rate in 40--56\% of trials, depending on its camp, whereas an agent 0.5 points above the post did so in only 12--27\%. Over 0.5--1.5 points, these readers moved 0.13 points (SE 0.02) further toward higher than lower posts, despite the disposition's bidirectional attraction.

\paragraph{Mirrored reading as a control.}
To control directional asymmetry, a fair coin decides whether an eligible post is read from the agent's position $x_i$ or the counterfactual position $2y_j - x_i$, at the same distance on the other side of the post. For a mirrored read, code reflects the reported change back onto the agent's position and clips to $[0,1]$. Posts at the reader's own rate, or with a mirror position outside the scale (including influencer posts), are read normally. This symmetrizes eligible responses in expectation while retaining LLM-generated changes, so the platform population is a controlled, symmetrized one rather than an unmodified LLM population.

\paragraph{Common tolerance and compatibility checks.}
Feeds use one estimated population threshold, not learned individual tolerances. A 20-round pilot of random and ultra-personalized feeds, which do not depend on $T$, located the mean response's attraction--repulsion boundary at $\hat T = 0.175$ (3.5 points). Main runs were conditional on ten compatibility checks (Appendix~\ref{app:gate}): distant repulsion in both directions, approximately balanced nearby attraction, and C1--C4, including adaptive reversal, in platform simulations resampling the measured updates.

\subsection{Experimental Scenarios}
\label{sec:llm-studies}

\paragraph{Study 1: feed comparison (C1--C4).}
The population starts from the bimodal initialization of Fig.~\ref{fig:init-bimodal}, with half of the agents drawn from $\mathcal{N}(0.8, 0.1^2)$ and half from $\mathcal{N}(0.3, 0.1^2)$, then clipped to $[0,1]$. The initial variance is approximately 0.074, with camps favoring about 20\% and 10\%. Six feeds run for 80 rounds: random (A1), ultra-personalized (A2, $E = 0.5^{20}$), tolerance-aware (A3, $E' = 100$), adaptive diversification alone (A4, $D = 0.15$), the combined design (A5), and tolerance-aware random (A6).

\paragraph{Study 2: diversification strength (C3).}
From a bimodal initialization, the combined design runs with $D \in \{0.05, 0.15, 0.35, 0.55, 0.95\}$ against A3 for 160 rounds, covering the regimes in Fig.~\ref{fig:sweep-D}.

\paragraph{Study 3: social pressure (C5).}
Following the scenarios in Fig.~\ref{fig:pressure}, a unimodal population encounters influencers at 4\% and 24\%, each entering a draw with probability 0.15. Their posts retain the exposure kernel but bypass the tolerance penalty. Under silence, each agent sees an opinion-climate cue: the nearest influencer's post, its own previous-round feed, or the previous round's average posted rate. It reports where most people stand ($\hat\mu_i$, corresponding to $x_{*,i}$), and code sets its posting probability to $S^{|x_i - \hat\mu_i|}$, with $S = 0.01$. Silent agents continue to read. Conditions (a)--(d) add pressure under the nearest-influencer cue: (a) personalized feeds ($E = 0.5^{10}$), (b) with influencers, (c) with influencers and silence, and (d) the combined design under both ($E' = 100$, $D = 0.5$). Conditions (e, f) and (g, h) repeat (c) and (d) with the other cues. Each condition runs for 30 rounds.

We report opinion variance and define reversal as a reduction of at least 0.05 (one percentage point) in the camp gap, the difference between the mean views of the two initial camps. Each arm and condition has three LLM runs (seeds 3026, 4026 and 5026), all using the pilot's $\hat T$; comparisons remain descriptive, and the reference runs characterize rule-based variability.

\subsection{Results by Arm and Additional Details}
\label{app:llm-outcomes}
\label{app:results}

Table~\ref{tab:llm} reports the change in camp gap, the final variance and the number of reversing runs and reference seeds for every arm of Studies~1 and~3. The three runs, with Step~1 and the pilot, required about 792,000 LLM calls at a total cost of about US\$52; no call in the three runs failed. In Study~1, agents starting in the outer third moved 2.3 percentage points toward the center on average (1.5--3.0 across runs), and one tolerance-aware run narrowly crossed the reversal threshold ($-0.051$). In Study~2, $D = 0.05$ narrowed the gap by 0.07--0.11 within about 65 rounds, but the gap partly reopened as the upper camp drifted upward ($-0.040$ at round 160; one run stayed reversed); $D = 0.55$ reversed polarization in one run.

Under random exposure, half of displayed posts lay beyond $\hat T$. The combined feed ended 0.152 below the tolerance-aware random feed on average (0.084--0.196) while displaying fewer posts across the midpoint (0.2\% versus 1.2\%). The approximately fixed LLM step sizes offer one explanation for differences from distance-scaled ARM responses.

Under silence, agents posted in 22--80\% of rounds. The combined feed's variance advantage was 0.008--0.020 on average across cues (0.001--0.033 in single runs), versus 0.029--0.057 in the reference. Personalized-feed variance was also lower on the LLM platform (0.045 versus 0.119 in condition a). Adding influencers did not raise variance over the 30-round horizon in any run.

\begin{table}[b]
\centering
\caption{Change in camp gap ($\Delta$gap) and final variance (Var.): LLM means over three runs (Rev.: runs reversing polarization) and rule-based reference means over 10 seeds (Rev.: seeds reversing polarization). $\hat T = 0.175$.}
\label{tab:llm}
\footnotesize
\setlength{\tabcolsep}{4.2pt}
\begin{adjustbox}{max width=\linewidth}
\begin{tabular}{@{}lrrrrrr@{}}
\toprule
& \multicolumn{3}{c}{LLM, three runs} & \multicolumn{3}{c}{Rule-based reference} \\
\cmidrule(lr){2-4}\cmidrule(l){5-7}
Arm & $\Delta$gap & Rev. & Var. & $\Delta$gap & Rev. & Var. \\
\midrule
\multicolumn{7}{@{}l}{\emph{Study 1, 80 rounds}} \\
A1 random & $+0.249$ & 0 & 0.235 & $+0.440$ & 0 & 0.250 \\
A2 ultra-personalized & $-0.013$ & 0 & 0.064 & $+0.024$ & 0 & 0.077 \\
A3 tolerance-aware & $-0.019$ & 1 & 0.061 & $-0.003$ & 0 & 0.066 \\
A4 adaptive only & $+0.109$ & 0 & 0.105 & $+0.126$ & 0 & 0.127 \\
A5 combined & $\mathbf{-0.113}$ & 3 & 0.041 & $-0.063$ & 7 & 0.052 \\
A6 tol.-aware random & $+0.039$ & 0 & 0.077 & $+0.280$ & 0 & 0.159 \\
\midrule
\multicolumn{7}{@{}l}{\emph{Study 2, 160 rounds}} \\
A3 (no $D$) & $+0.012$ & 0 & 0.069 & $-0.003$ & 0 & 0.066 \\
$D = 0.05$ & $-0.040$ & 1 & 0.058 & $-0.088$ & 9 & 0.046 \\
$D = 0.15$ (A5) & $\mathbf{-0.127}$ & 3 & 0.039 & $-0.076$ & 7 & 0.050 \\
$D = 0.35$ & $\mathbf{-0.103}$ & 3 & 0.043 & $-0.009$ & 3 & 0.066 \\
$D = 0.55$ & $-0.041$ & 1 & 0.056 & $+0.120$ & 0 & 0.102 \\
$D = 0.95$ & $+0.091$ & 0 & 0.094 & $+0.390$ & 0 & 0.209 \\
\bottomrule
\end{tabular}
\end{adjustbox}
\end{table}

\subsection{Rule Fits and Pilot Checks}
\label{app:gate}

We fitted six update rules to the 486 trials in which readers held rates on the grid, using five-fold cross-validation grouped by persona identifier. Their cross-validated log-likelihoods were 923.1 for fixed steps, 910.4 for attraction--repulsion with separate attraction and repulsion rates, 895.8 for attraction--repulsion with one rate, 814.1 for linear assimilation, 781.1 for bounded confidence and 750.7 for no change. Table~\ref{tab:gate} reports ten compatibility checks, all of which passed. Outcome checks use platform simulations resampling measured updates by distance and position, rather than new LLM runs. Because the gates require the expected qualitative policy outcomes, passing them establishes compatibility of the selected response process; it does not independently corroborate C1--C4.

The pilot estimates a common $\hat T$ from the sign change in the mean update toward a post, interpolating between distance bins and rounding to 0.025. It does not estimate each persona's tolerance. The code tests two candidate disposition wordings in order and uses the first passing all gates. The first, W4, passed, so the second was not run; the disposition quoted below is W4.

\begin{table}[t]
\caption{Compatibility gates, on the 0--1 scale (one point is 0.05). $\Delta$gap: change in camp gap in platform simulations resampling measured updates (10 seeds; brackets give the share reversing polarization). These gates condition the main runs on the expected qualitative outcomes.}
\label{tab:gate}
\footnotesize
\begin{adjustbox}{max width=\linewidth}
\begin{tabular}{@{}p{0.56\linewidth}p{0.38\linewidth}@{}}
\toprule
Check & Value \\
\midrule
\multicolumn{2}{@{}l}{\emph{Step 1: single-post trials}} \\
A rule that repels fits best & fixed steps (see text) \\
Its edge lies between 0.2 and 0.3 & 0.25 [0.225, 0.275] \\
Higher and lower posts pull alike (below 0.01) & 0.0024 (SE 0.0036) \\
Posts beyond the edge push away, both ways & $-0.037$ and $-0.048$ \\
Its moves show C1--C4 in our model & $\Delta$gap A1 $+0.44$, A2 $+0.00$, A5 $-0.12$ \\
\midrule
\multicolumn{2}{@{}l}{\emph{Pilot: 20 rounds of A1 and A2, 7,071 reads; edge $\hat T = 0.175$}} \\
Posts beyond the edge push away, both ways & $-0.024$ and $-0.040$ \\
Close posts pull up and down alike (below 0.003) & $-0.0003$, $-0.0006$ (SE 0.0008) \\
Its moves show C1--C4 in our model (80 rounds) & $\Delta$gap A1 $+0.40$, A2 $+0.00$, A5 $-0.07$ [0.6], A6 $+0.05$ \\
A5 reverses, A3 does not (160 rounds) & $\Delta$gap A5 $-0.08$ [0.7], A3 $-0.00$ \\
$D = 0.95$ does not reverse, ends above A5 & $\Delta$gap $+0.13$ \\
\bottomrule
\end{tabular}
\end{adjustbox}
\end{table}

\subsection{Prompts}
\label{app:prompts}

All prompts are in English, and every agent call uses the system message ``You are taking part in a study of how people use social media. Stay in character as the person described below. Answer in JSON only.'' The user message begins with a persona block, shown here as an agent receives it when reading a post:

\begin{prompttext}
You are Uma Johnson, 30, a factory technician in Nova, a fictional U.S. state, where you live in a suburb of the state's largest city. You moved to Nova five years ago for work. The topic is a question before Nova's state legislature: what share of their wages should workers be required to pay into a public retirement fund run by the state? Lawmakers can set any rate from 4\% to 24\% of wages. A higher rate means bigger pensions later but less take-home pay now; a lower rate means more take-home pay now but smaller pensions later. You have an opinion on this issue, but it is not central to who you are. A good argument from someone whose view is close to yours, whether they want a slightly higher rate or a slightly lower one, can win you over. A post from someone whose view is more than a few points away from yours, in either direction, tends to strike you as one-sided, and it can push your own view further away from theirs.
\end{prompttext}

\noindent Personas combine 60 first names, 30 surnames, ages from 18 to 45, 20 occupations, 8 places and 10 backgrounds. Half of the agents have low involvement, as in this example, and the other half are told ``This issue is one of the few you care about most. It is tied to who you are.'' The disposition, given by the last two sentences, appears only in prompts that ask for the agent's view, when it reads a post and in the no-post control. The persona block is followed by the view line ``Your current view is that Nova should set the contribution at \slot{rate}\% of wages.'' (the mirror position under mirrored reading) and then by the task.

\begin{samepage}
\noindent\emph{Writing a post.}
\begin{prompttext}
Write a short social-media post (at most 50 words) giving your view on the share of wages that workers in Nova should be required to pay into the state retirement fund. Write it the way you would actually write it.\\
Reply only with JSON: \texttt{\{"post": "<text>"\}}
\end{prompttext}
\end{samepage}

\begin{samepage}
\noindent\emph{Reading a post} (Step~1 and the platform).
\begin{prompttext}
You come across this post from another user: ``\slot{post}''\\
Would you like this post? Then, after reading it, what contribution rate do you now think Nova should set (a number from 4 to 24)?\\
Reply only with JSON: \texttt{\{"like": true or false, "view": <number from 4 to 24>, "why": "<one short sentence>"\}}
\end{prompttext}
\end{samepage}

\begin{samepage}
\noindent\emph{No post} (Step~1 controls).
\begin{prompttext}
Take a moment to think about the topic again. Then, what contribution rate do you now think Nova should set (a number from 4 to 24)?\\
Reply only with JSON: \texttt{\{"view": <number from 4 to 24>, "why": "<one short sentence>"\}}
\end{prompttext}
\end{samepage}

\noindent\emph{Climate} (Study~3, under silence). One of three cues comes first: ``A popular account that many people here follow posted this today: \slot{post of the influencer on the agent's side of the middle}''; ``Posts you saw on this platform yesterday: 1.~\slot{post} 2.~\ldots''; or ``On this platform, the average rate favored in yesterday's posts was \slot{rate}\% of wages.'' Then:
\begin{prompttext}
What contribution rate do you think most people on this platform favor (a number from 4 to 24)?\\
Reply only with JSON: \texttt{\{"most\_people": <number from 4 to 24>\}}
\end{prompttext}

\noindent\emph{Rater} (Gemini 3.8 Flash, ten posts per call). The system message reads ``You read social-media posts about the share of wages that workers in Nova should be required to pay into the state retirement fund (any rate from 4\% to 24\% of wages). Answer in JSON only.''
\begin{prompttext}
For each post, give the contribution rate its author favors, as a number from 4 to 24 (\% of wages). Use 14 if the post takes no side.\\
Posts: 1.~\slot{post} 2.~\ldots\\
Reply only with JSON: \texttt{\{"rates": [<number>, ...]\}}
\end{prompttext}

\end{document}